\documentclass[leqno]{article}
\usepackage{epsfig}
\usepackage{graphics,graphicx}
 \usepackage{amsmath}
 \usepackage{amssymb}
\usepackage{amsthm}
\usepackage{hyperref}

\DeclareRobustCommand{\qed}{%
  \ifmmode 
  \else \leavevmode\unskip\penalty9999 \hbox{}\nobreak\hfill
  \fi
\quad\hbox{\qedsymbol}}
\newcommand{\mathbold}[1]{\mbox{\boldmath $#1$}}
\newcommand{\rz}{\mathbb R}

\newcommand{\pr}{\mathbold P}

\begin{document}
\begin{center}
{\large Statistical Analysis of the Ricker Model}

\quad\\
Laurie Davies\\
{\it Faculty of Mathematics\\
University of Duisburg-Essen, 45117 Essen, Federal Republic of
Germany\\
e-mail:laurie.davies@uni-due.de}
\end{center}
\quad\\

\begin{abstract}
The Ricker model was introduced in the context of managing fishing
stocks. It is a discrete non-linear iterative model given by
$N(t+1)=rN(t)\exp(-N(t))$ where $N(t)$ is the population at time
$t$. The model treated in this paper includes a random component
$N(t+1)=rN(t)\exp(-N(t)+\varepsilon(t+1))$ and what is observed at
time $t$ is a Poisson random variable with parameter $\varphi
N(t)$. Such a model has been analysed using `synthetic likelihood' and
ABC (Approximate Bayesian Computation). In contrast this paper takes a
non-likelihood approach and treats the model in a consistent manner as
an approximation. The goal is to specify those parameter values if any
which are consistent with the data.
\end{abstract}

\section{Introduction}
The stochastic Ricker model is defined by
\begin{equation} \label{equ:pois}
Y(t)= \text{rpois}(\varphi N(t))
\end{equation}
where $\text{rpois}(\lambda)$ is a Poisson random variable with
parameter $\lambda$, $\varphi$ is a scale parameter and $N(t)$ is a
stochastic process defined by
\begin{equation} \label{equ:stoch_proc}
\log(N(t+1))=\log(r)+\log(N(t))-N(t)+\sigma \varepsilon(t+1)
\end{equation}
where $\varepsilon$ is standard Gaussian noise and $r$ and $\sigma$
are further parameters. There are three parameters
$\theta=(r,\sigma,\varphi)$ in all.

Given data $y(t), t=1,\ldots,n,$ the aim of this paper is to specify
those parameter values $\theta$ if any for which the stochastic Ricker
process $Y_{\theta}(t)$ is an adequate approximation to the data. This
is done by calculating several statistics associated with the data and
using simulations to determine their typical values under the model. If the
values from the data $y(t)$ belong to the set of typical values for
the parameter $\theta$ then this model is an adequate approximation to
the data. The word `typical' is made precise by requiring that for
data $Y_{\theta}(t)$ generated under the model the associated values
are typical with probability $\alpha$. What is meant by an `adequate
approximation' is defined by the choice of the statistics. 
 
More formally, given a model $P_{\theta}$ the statistician defines a subset
$E_{n,\theta}$ of $\rz^n$ such that for data ${\mathbold
  Y}_{n,\theta}=(Y_{1,\theta},\ldots,Y_{n,\theta})$
\begin{equation} \label{equ:approx_reg_1}
\pr({\mathbold Y}_{n,\theta} \in E_{n,\theta})=\alpha.
\end{equation}
The probability $\alpha$ defines `typical' and the set $E_{n,\theta}$
defines `look like'. The subset $E_{n,\theta}$ is defined through a
finite number of statistics $T_{j,n,\theta}, j=1,\ldots,k$ and their
typical values $(q_{l,j,\theta}(\alpha_j),q_{u,j,\theta}(\alpha_j))
j=1,\ldots,k$ satisfying
\begin{equation} \label{equ:inequ_1}
\pr(q_{l,j,\theta}(\alpha_j)\le T_{j,n,\theta}({\mathbold
Y}_{n,\theta})\le q_{u,j,\theta}(\alpha_j))=\alpha_j, j=1,\ldots,k.
\end{equation}
where 
\begin{equation} \label{equ:sum_alpha_1}
\sum_{j=1}^k(1-\alpha_j)=1-\alpha.
\end{equation}
If the $\alpha_j$ satisfy (\ref{equ:sum_alpha_1}) then
the equals sign $=$ in (\ref{equ:approx_reg_1}) must be replaced by
$\ge$. If equality in (\ref{equ:approx_reg_1}) is required the
$\alpha$ in (\ref{equ:sum_alpha_1})  can be replaced by 
\begin{equation} \label{equ:sum_alpha_2}
\sum_{j=1}^k(1-\alpha_j)=1-{\tilde \alpha}.
\end{equation}
where ${\tilde   \alpha}<\alpha$ chosen such that
(\ref{equ:approx_reg_1}) holds. This can be done by using simulations
to obtain the actual covering probability $\alpha^*$ in
(\ref{equ:approx_reg_1}) and then putting ${\tilde
  \alpha}=2\alpha-\alpha^*$.

For a given data set ${\mathbold y}_n$ not necessarily generated under
the model the approximation region ${\mathcal A}({\mathbold
  y}_n,\alpha,\Theta)$ is defined by
\begin{equation}
{\mathcal A}({\mathbold
  y}_n,\alpha,\Theta)=\left\{\theta: {\mathbold y}_n\in E_{n,\theta}\right\}.
\end{equation}
This is not a confidence interval and may well be empty. The approach
to statistics expounded in \cite{DAV14} is based on this simple idea.

There is no automatic choice of the statistics. It will 
depend on the subject matter, the model and also on practical
considerations such as computability. In the present paper for the
Ricker model five statistics will be used. They can of course be
subject to criticism and there may well be better ones. It is often
difficult to capture the essence so to speak of a data set by
specifying a number of numerical values. For the difficulties of
doing this for long range financial data see \cite{DAVKRA16}.
Sometimes eye-balling may be the best option as one can recognize an
elephant although a numerical description is difficult. For further
examples and discussions see \cite{NEYSCOSHA53},
\cite{NEYSCOSHA54},\cite{BUJETAL09} and pages 31-32 and page 112 of
\cite{HUB11}.

The approach taken here contrasts with the likelihood approach of 
\cite{WOO10} and \cite{GUTCOR15}, this latter in spite of its title. 

\section{Choosing the statistics}
\subsection{Typical sample values}
\begin{figure}
\begin{center}
\includegraphics[width=4cm,height=12cm,angle=270]{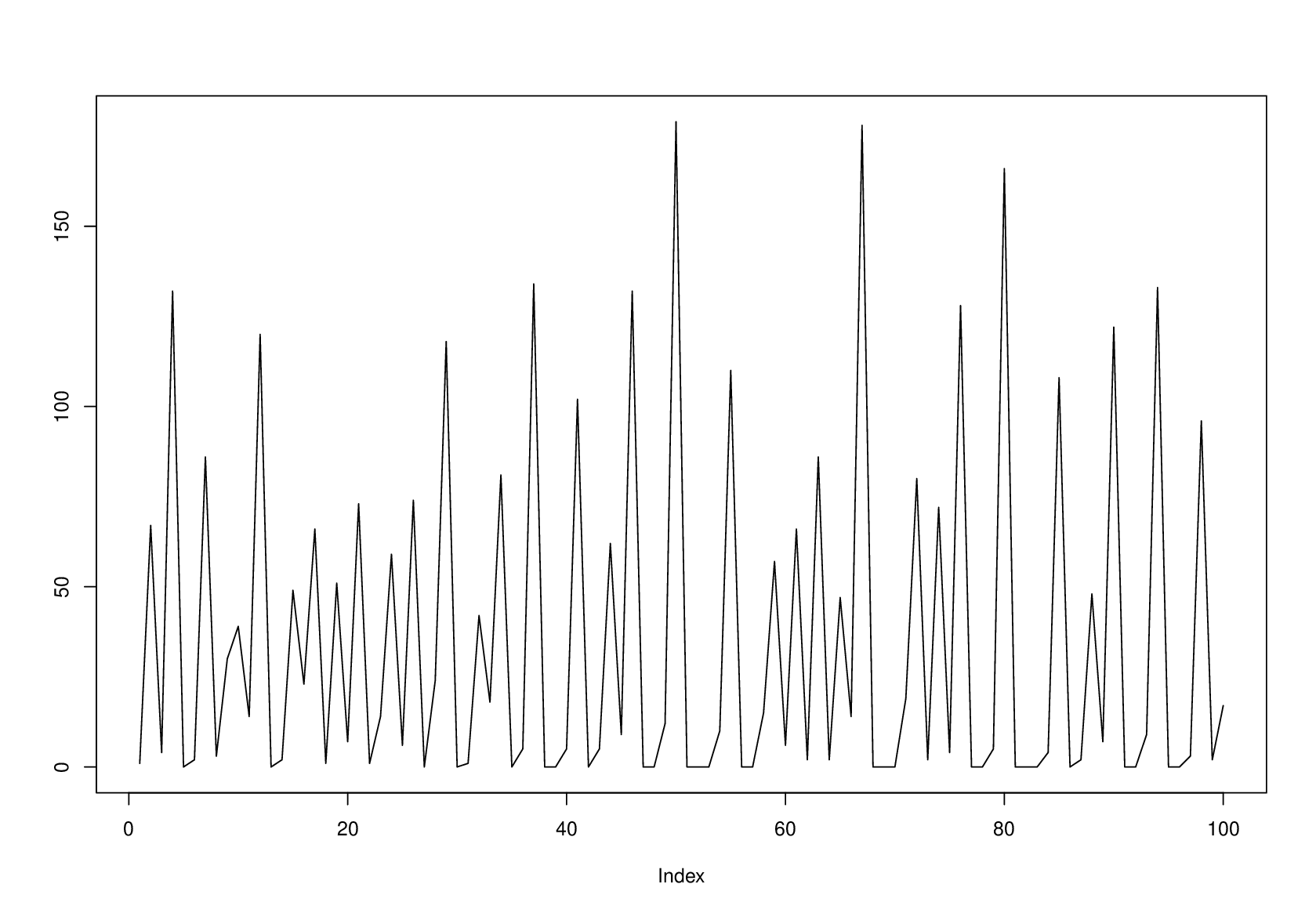}
\includegraphics[width=4cm,height=12cm,angle=270]{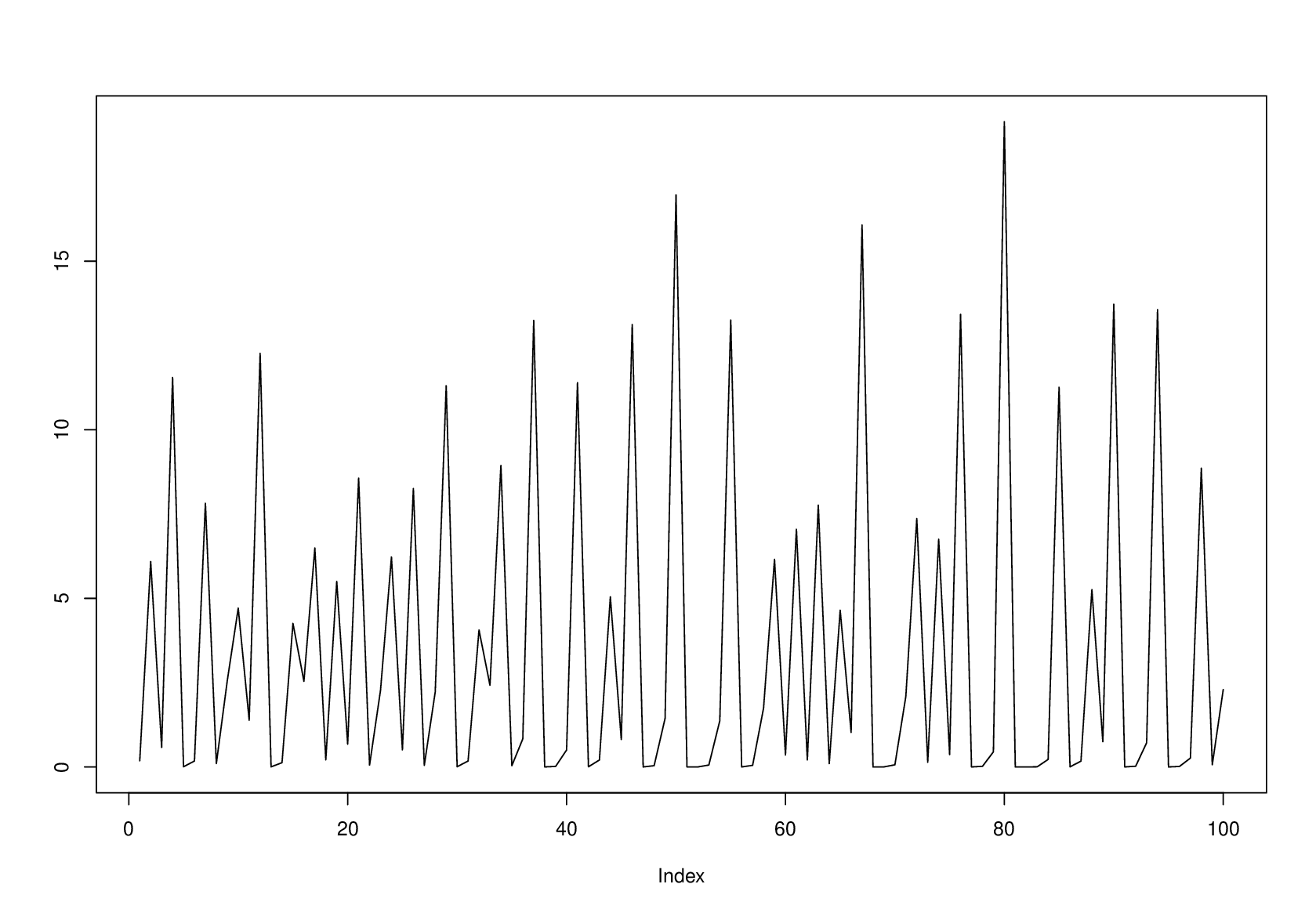}
\includegraphics[width=4cm,height=12cm,angle=270]{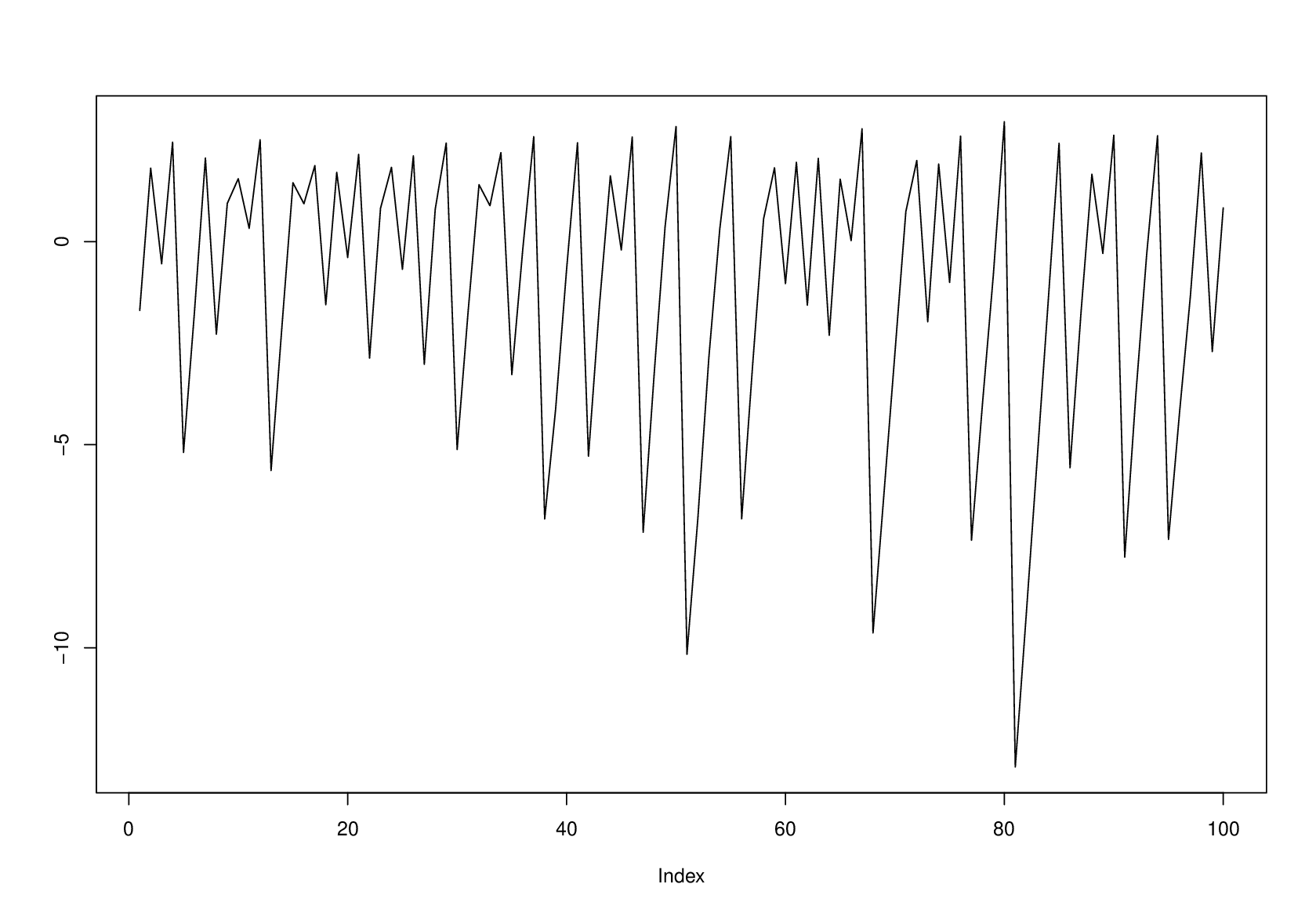}
\end{center}
\caption{The processes $Y(t), N(t)$ and $\log(N(t))$ from top to
  bottom. \label{fig:one}} 
\end{figure}

The top panel of Figure~\ref{fig:one} shows a realization of size
$n=100$ of such a process with 
\begin{equation}
\theta=(\exp(3.6),0.3,10),
\end{equation}
the centre panel shows the corresponding $N(t)$ process and the bottom
panel the $\log(N(t))$ process.

We concentrate initially on the typical values of the $\log(N(t))$
process, that is on the order statistics. This ignores the dynamics of
$\log(N(t))$ which will be treated in the 
Section~\ref{sec:dynamics}. The idea is to compare the values of the
data $y(t)$ with those of $\varphi \exp(f(t))$ using the Kolmogorov
metric where the $f(t)$ are the expected values of the order
statistics of $\log N(t)$. The function $f(t)$ will be approximated by
a parametric function $f_{n,r,\sigma}(t)$ involving 20 parameters
obtained by a linear regression. There may be better ways of
calculating a simple approximation but the one given is sufficiently
accurate for the comparison. It will be described in the
Section~\ref{sec:tables}. The parameters will be stored for all
$(r,\sigma)$ on a suitably fine grid.

\begin{figure}
\begin{center}
\includegraphics[width=4cm,height=12cm,angle=270]{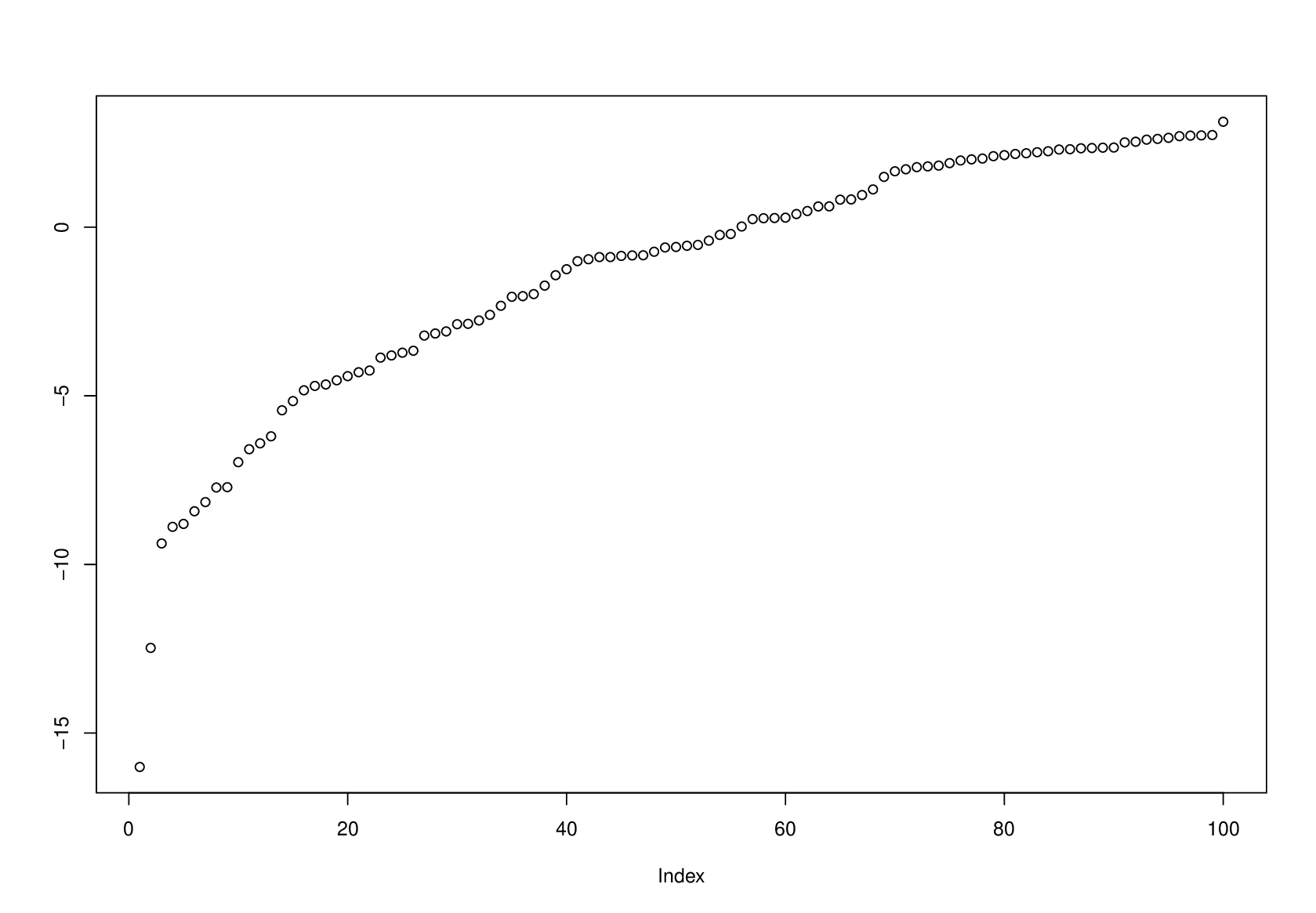}
\includegraphics[width=4cm,height=12cm,angle=270]{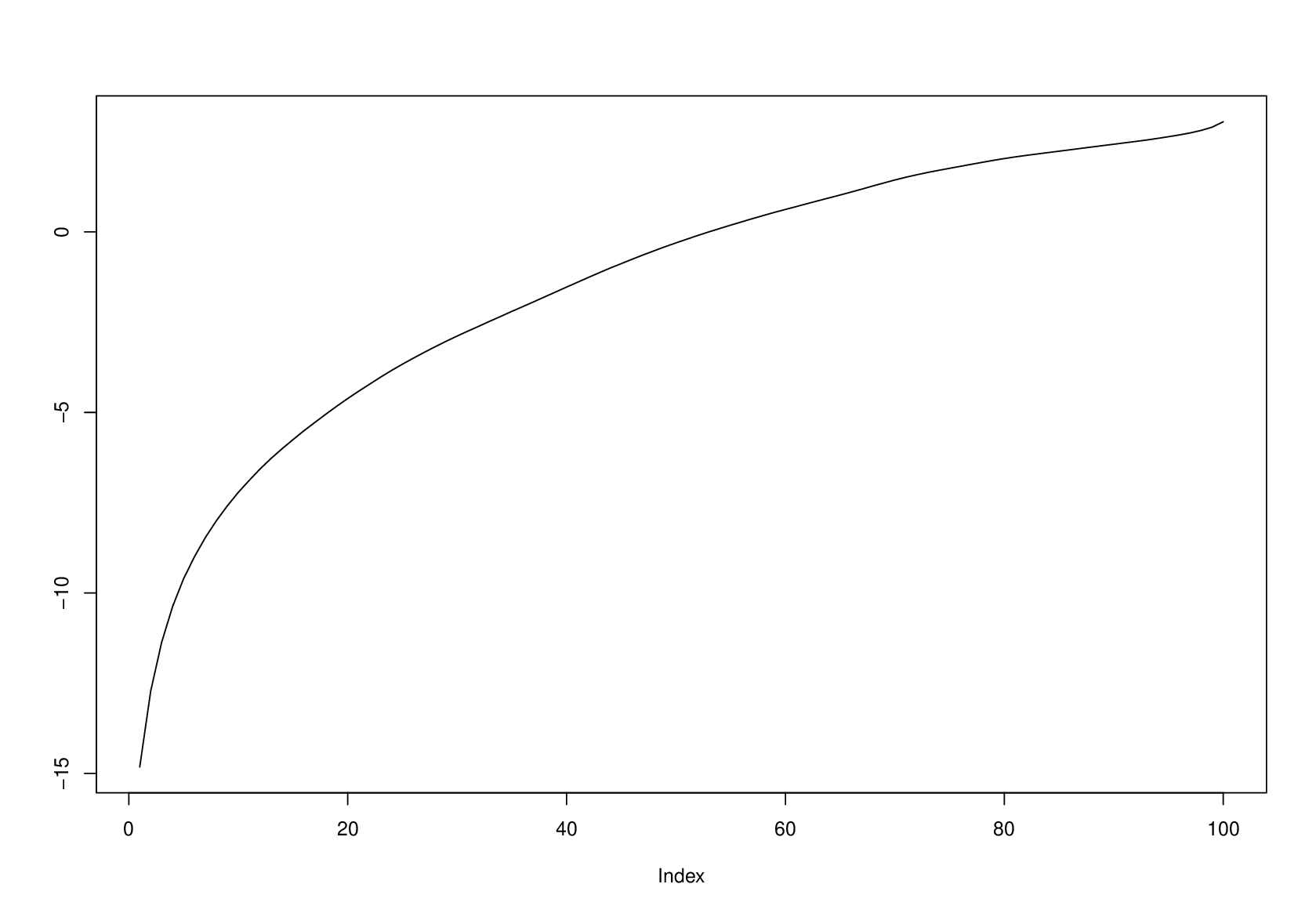}
\includegraphics[width=4cm,height=12cm,angle=270]{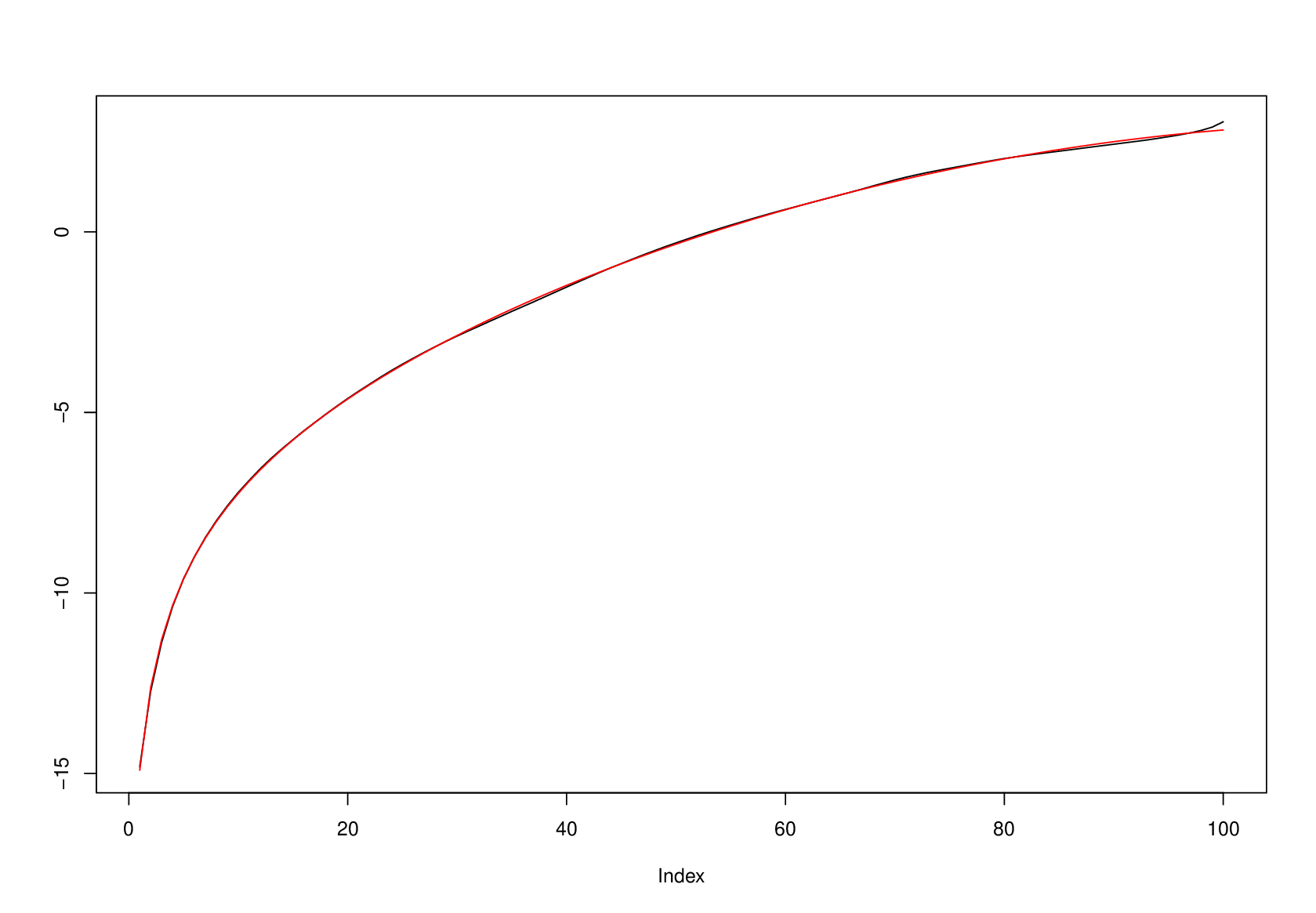}
\end{center}
\caption{Top panel: The ordered values of the bottom panel of
  Figure~\ref{fig:one}. Centre panel:  the means of the
  ordered values of 500 simulated samples. Bottom panel: the parametric
  approximation $f_{n,r,\sigma}(t)$  to the centre panel in
  red.\label{fig:three}}  
\end{figure}

The top panel of Figure~\ref{fig:three} shows the
order statistics of the bottom panel of Figure~\ref{fig:one}.
The centre panel shows the mean of the order statistics calculated by
simulating 500 samples of size 100. The bottom panel shows the
parametric approximation $f_{n,r,\sigma}(t)$ in red.

The order statistics of the process $N(t)$ can be approximated by
$\exp(f_{n,r,\sigma}(t))$. Given $\varphi$ the order statistics of
$Y(t)=\text{rpois}(\phi N(t))$ for a sample of size $n$ may be
approximated by the function $\varphi\exp(f_{n,r,\sigma})$. As the
$Y(t)$ are integers and the measure of approximation to be used is the
Kolmogorov metric the values of $\varphi\exp(f_{n,r,\sigma})$ are
replaced by the nearest integer $\text{int}(\varphi\exp(f_{n,r,\sigma}))$.

\begin{figure}
\begin{center}
\includegraphics[width=4cm,height=12cm,angle=270]{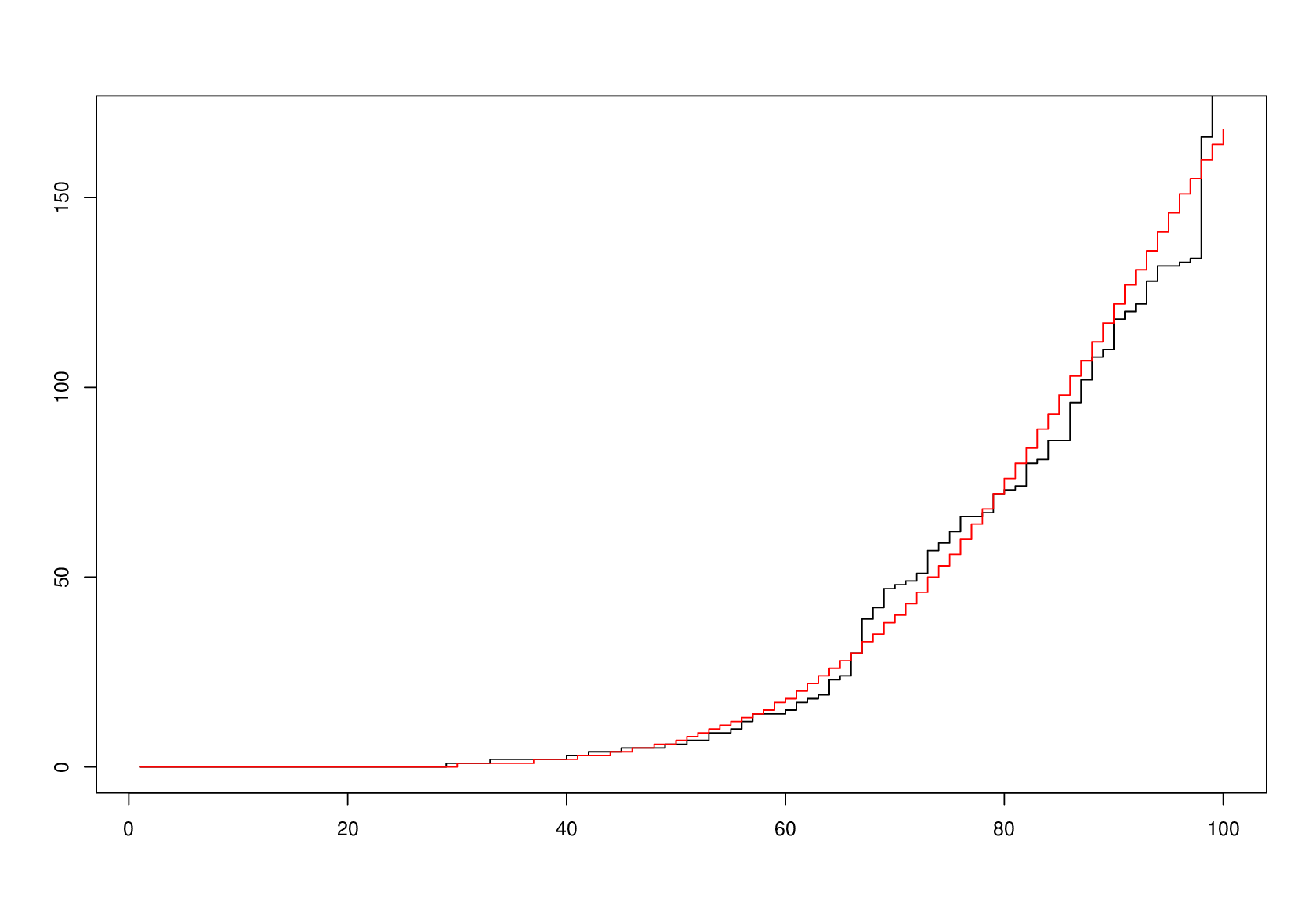}
\includegraphics[width=4cm,height=12cm,angle=270]{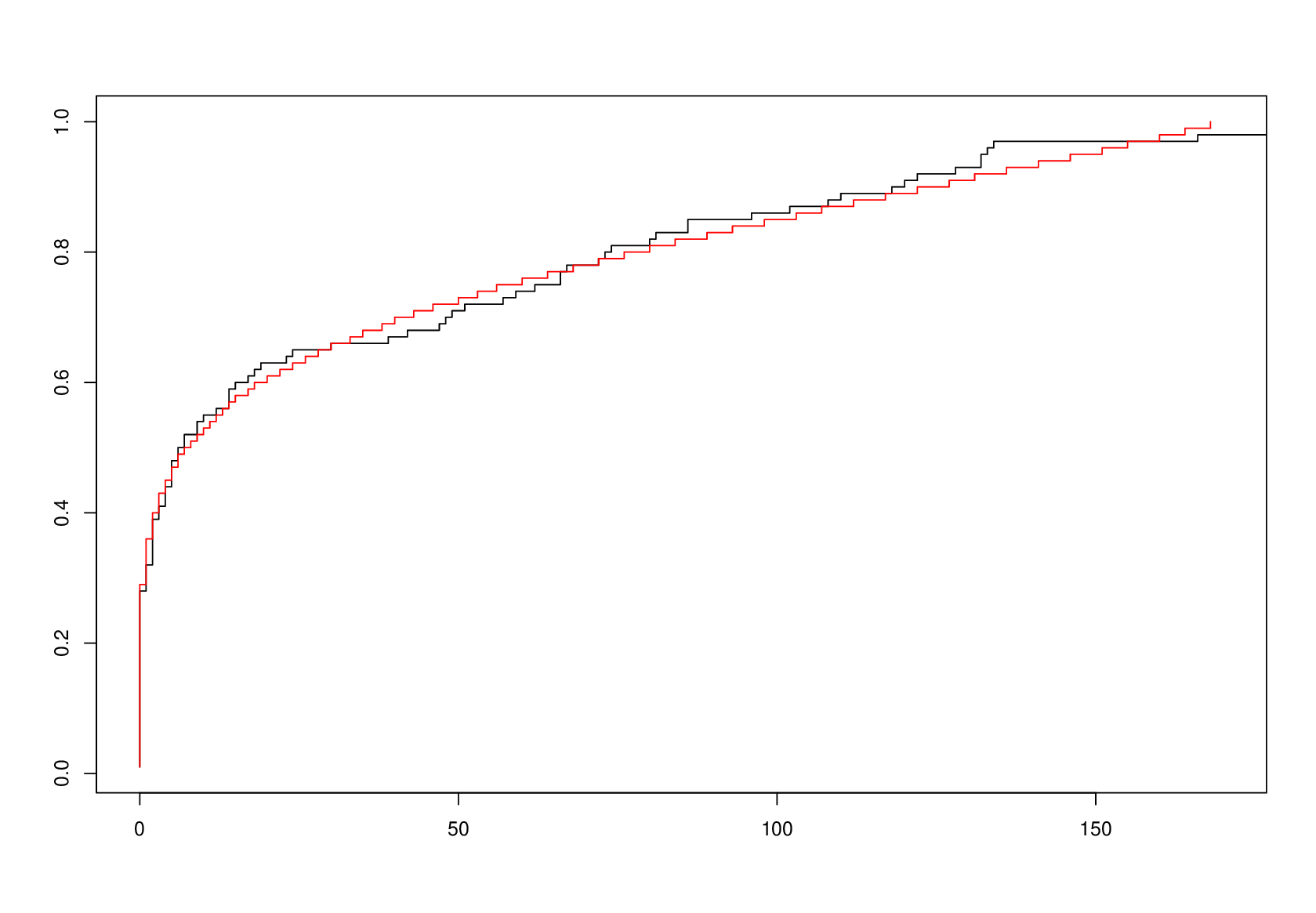}
\end{center}
\caption{Upper panel: the order statistics of the first 100 values of
  the upper panel of  Figure~\ref{fig:one}  in black and the
  parametric approximation
  $\text{int}(10\exp(f_{100,\exp(3.6),0.3}))$ in red. Lower panel: the
  distribution functions. The Kolmogorov distance is 0.06.\label{fig:five}} 
\end{figure}

The upper panel of Figure~\ref{fig:five}  shows the ordered sample of
the upper panel of Figure~\ref{fig:one} in black and the function 
$\text{int}(10\exp(f_{100,\exp(3.6),0.3}))$ in red. The lower panel shows
the distribution functions. The Kolmogorov distance is
0.06. 
Based on 2000 simulations the  0.95-quantile of the Kolmogorov
distance with for data generated with $\theta=(\exp(3.6),0.3,10)$ is
0.11. Thus not surprisingly the data of Figure~\ref{fig:five} are
consistent, in the sense of the Kolmogorov metric, with
$\theta=(\exp(3.6),0.3,10)$ used to generate them.  

\begin{figure}
\begin{center}
\includegraphics[width=4cm,height=12cm,angle=270]{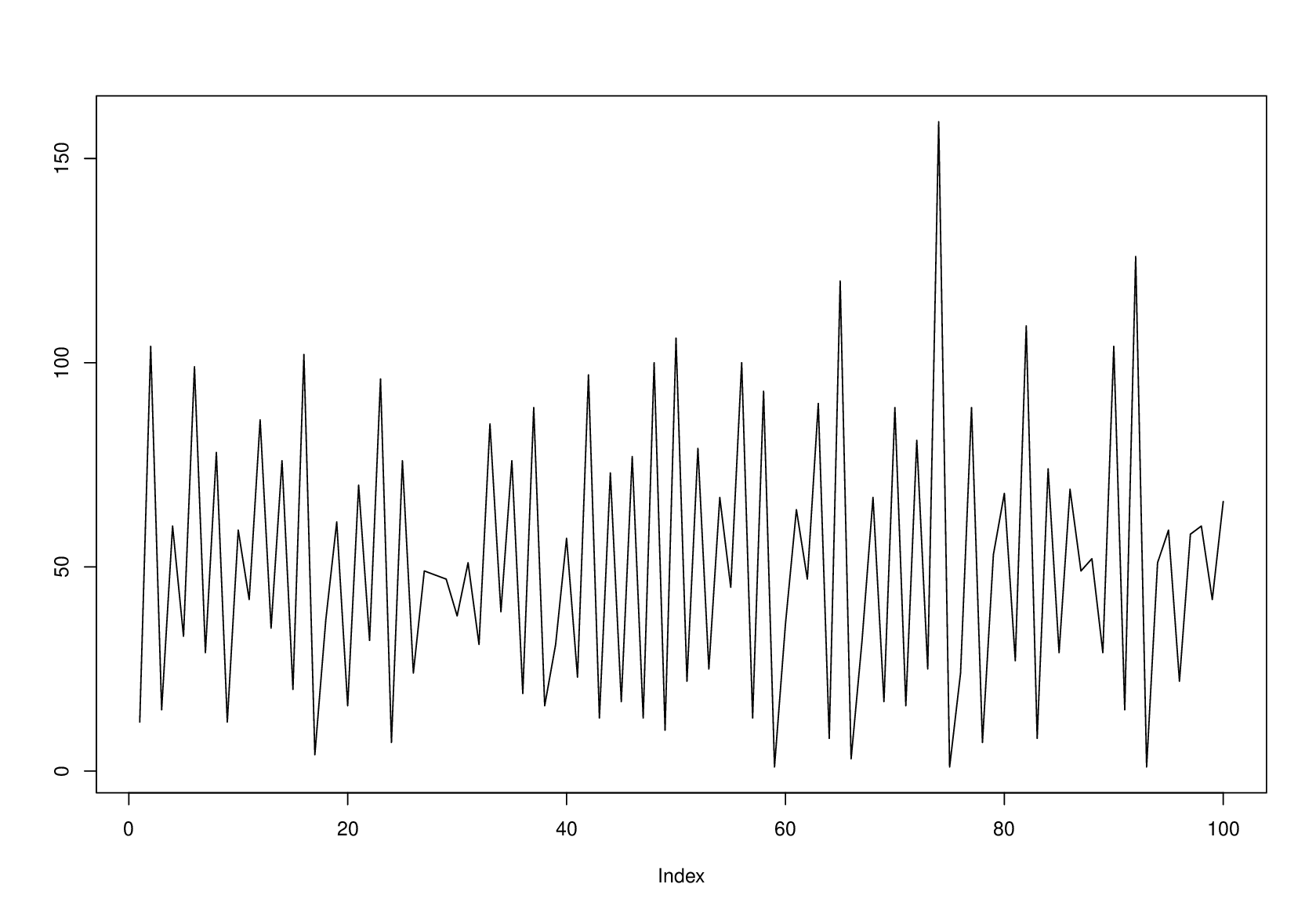}
\includegraphics[width=4cm,height=12cm,angle=270]{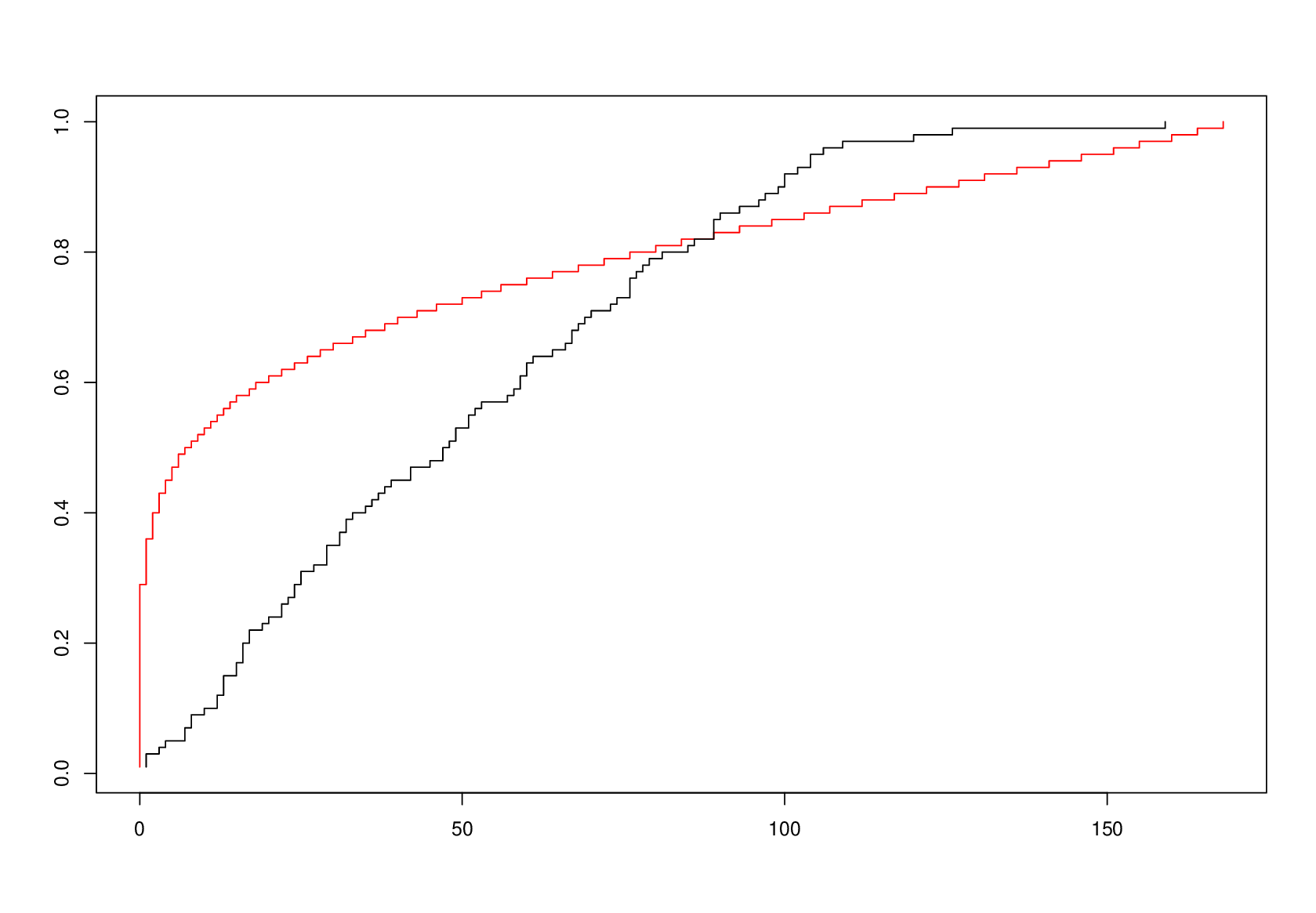}
\end{center}
\caption{Upper panel: a sample of size $n=100$ generated with
  $\theta=(\exp(2.6),0.3,20)$. Lower panel: the distribution function of the
  data and that of the parametric approximation
  $\text{int}(10\exp(f_{100,\exp(3.6),0.3}))$ in red.  The Kolmogorov
  distance is 0.44.\label{fig:six}}
\end{figure}

The upper panel of Figure~\ref{fig:six} shows a data of size $n=100$
generated with $\theta=(\exp(2.6),0.3,10)$. The lower panel shows its
distribution function in black and that of
$\text{int}(10\exp(f_{100,\exp(3.6),0.3}))$ in red. The Kolmogorov
distance is 0.37 which far exceeds the 0.95-quantile of 
0.11 for data generated with $\theta=(\exp(3.6),0.3,10)$. The
conclusion is that the data are not consistent with the model
$\theta=(\exp(3.6),0.3,10)$.

The above requires a value for $\varphi$ which may be obtained from
the data as follows.  Given the $N(t)$ the $Y(t)$ are defined by 
$Y(t)=\text{rpois}(\varphi N(t))$ so that
\begin{equation} \label{equ:varphi_approx_1}
\sum_{t=1}^n Y(t) \stackrel{D}{=}\text{rpois}\Big(\varphi\sum_{t=1}^n
N(t)\Big).
\end{equation}
Given $\beta$ a $\beta$-approximation interval for $\varphi
\sum_{i=1}^n N(t)$ is given by $[\lambda_l,\lambda_u]$ where
\begin{equation} \label{equ:varphi_approx_2}
\text{ppois}\Big(\sum_{t=1}^nY(t),\lambda_l\Big)=(1+\beta)/2, \quad
\text{ppois}\Big(\sum_{t=1}^nY(t),\lambda_u\Big)=(1-\beta)/2.
\end{equation}
which translates into the approximation interval
\begin{equation} \label{equ:varphi_approx_3}
\left[\,\frac{\lambda_l}{\sum_{t=1}^n N(t)},\frac{\lambda_u}{\sum_{t=1}^n N(t)}\,\right]
\end{equation}
for $\varphi$.

A $\beta$-approximation interval for $\sum_{i=1}^nN(t)$
is given by the $(1-\beta)/2$ and $(1+\beta)/2$ quantiles
$q_{l,N}(\beta)$ and $q_{u,N}(\beta)$ respectively. These can
be obtained by simulations and saved together with the 20
parameter values required for $f_{n,r,\sigma}(t)$ giving 22 values in
all. The default value of $\beta$ is $\beta=0.99$. The final
approximation interval for $\varphi$ is
\begin{equation} \label{equ:varphi_approx_4}
\left[\,\frac{\lambda_l}{q_{u,N}(\beta)},\frac{\lambda_u}{q_{l,N}(\beta)}\,\right].
\end{equation}

As an example for the data of Figure~\ref{fig:one} we have
$\sum_{t=1}^nY(t)=3473$. For $\beta=0.99$  it follows from
(\ref{equ:varphi_approx_2}) that $\lambda_l=3323$ and
$\lambda_u=3627$. For $(r,\sigma)=(\exp(3.6),0.3)$  simulations give 
$q_{l,N}(0.95)=342.0$ and $q_{u,N}(0.95)=378.2$. The final
approximation region for $\varphi$ is
$[3323/378.2,\,3627/342.0]=[8.79,\,10.61]$ 

Given such an interval a grid can be placed on it and the quantiles of
the Kolmogorov metric obtained through simulations. These can be
compared with the actual Kolmogorov distance of the data from
$\text{int}(\varphi\exp(f_{n,r,\sigma}))$.

\subsection{The dynamics of the process}\label{sec:dynamics}
The Kolmogorov distance depends only on the empirical distribution
functions and takes no account of the dynamics of the process. This
will be done by mimicking the dynamics of the $\log N(t)$ process by
regressing $\log(y(t)/\varphi+\delta)$ on
$\log(y(t-1)/\varphi+\delta)$ and $y(t-1)/\varphi$ for a choice of
$\varphi$ and with default choice $\delta=0.01$. 

\begin{figure}
\begin{center}
\includegraphics[width=4cm,height=12cm,angle=270]{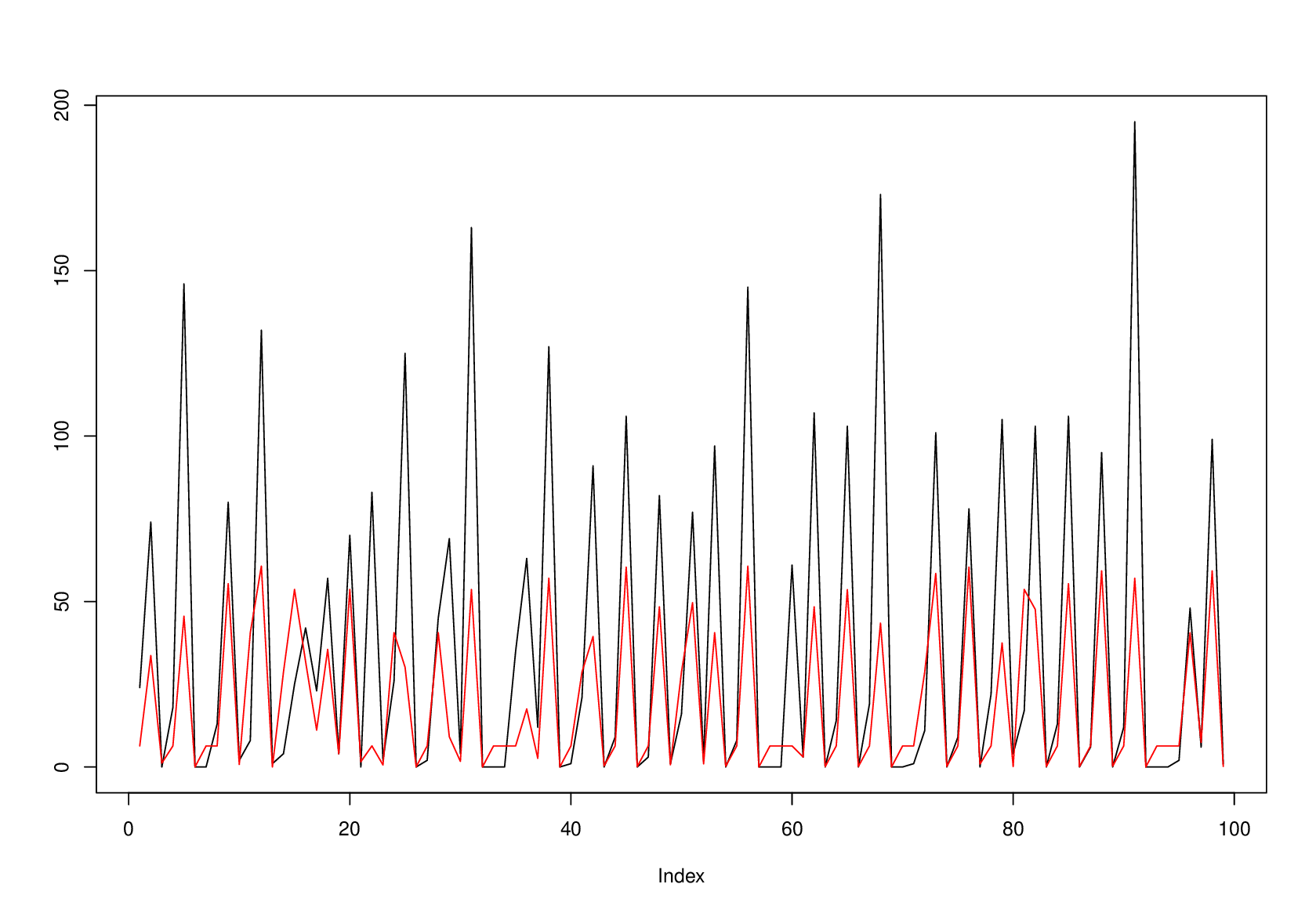}
\includegraphics[width=4cm,height=12cm,angle=270]{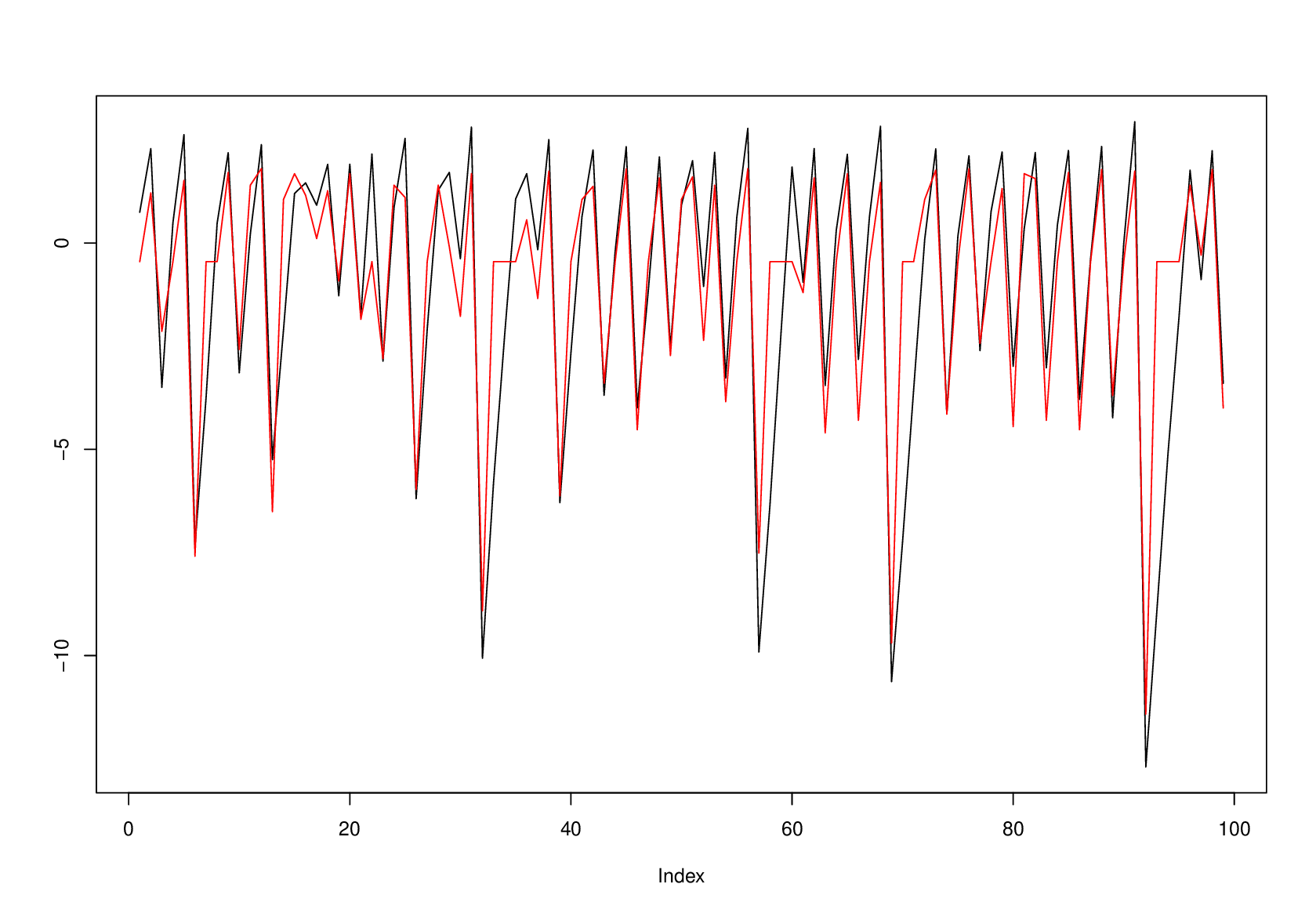}
\end{center}
\caption{Upper panel: A data set generated with
  $\theta=(\exp(3.6),0.3,10)$ together with the forecast for $Y(t)$
  based on $Y(t-1)$. Lower panel: the same for the $N(t)$ process. \label{fig:reg_1}} 
\end{figure}

The upper panel of Figure~\ref{fig:reg_1}  shows a data set $Y(t)$ of size $n=100$
(black) generated with $\theta=(\exp(3.6),0.3,10)$ together with the forecast (red)
${\tilde Y}(t)$ of $Y(t)$ based on $Y(t-1)$;
\begin{equation}
{\tilde
  Y}(t)=\varphi\exp(\beta_1))(Y(t-1)/\varphi+\delta)^{\beta_2}
\exp(\beta_3 (Y(t-1)/\varphi+\delta))   
\end{equation}
where the $\beta_j$ with $\beta=(2.61,0.66,-0.82)$ are the regression
coefficients.  the lower panel does the same for the $N(t)$ process.
The R output for the regression is
\begin{equation}
\left\{
\begin{array}{ccccc}
   &\text{Estimate}&\text{ Std. Error}&t&\text{value } Pr(>|t|)\\    
\beta_1&            2.60673 &   0.43069 & 6.052& 2.76e-08 ***\\
\beta_2&              0.66255 &   0.09178 &  7.219& 1.23e-10 ***\\
\beta_3& -0.81973&    0.06997& -11.716& < 2e-16 ***\\
\end{array}
\right.
\end{equation}
with residual standard error: 2.381.

There are four statistics, the three values of the
coefficients and the standard deviation of the residuals. Given a
parameter $\theta$ simulations will give lower and upper bounds for
these four values and these can be compared with the values 
obtained from the data at hand.

\subsection{Accounting}
Given a value of $\alpha$ in (\ref{equ:approx_reg_1}) the statistician
has $1-\alpha$ of probability to spend. In all there are six
statistics to be paid for: (i) the bounds for $\varphi$ as in
(\ref{equ:varphi_approx_4}), (ii) the Kolmogorov distance, (iii)-(v)
bounds for the coefficients in the regression and (vi) for the
standard deviation of the residuals.  

The bounds for $\varphi$ differs in their construction from the others in
using the lower and upper quantiles for $\sum_{t=1}^{100} N(t)$ in 
 (\ref{equ:varphi_approx_4}). The result is that the covering probability
 is almost one, much higher than the 0.98 would indicate. In a
 simulation study with 5000 simulations and $\theta=(\exp(3.6),0.3,10)$ the
interval included the value of $\varphi=10$ in all but three of the
the simulations, The actual cost of the interval for $\varphi$ is
therefore very small and will be ignored.

Of the remaining five statistics the simplest choice is to spend an
equal amount $(1-\alpha)/5$ on each, $\alpha_j=(4+\alpha)/5,
j=1,\ldots,5,$  as in (\ref{equ:sum_alpha_1}). In the case of the Kolmogorov
distance it seems reasonable just to use the upper bound. This can
also be argued for the standard deviation of the residuals but less
convincingly. In all other cases lower and upper bounds are appropriate
and treating these equally leave $(1-\alpha)/10$ to be spent on each.

As mentioned after (\ref{equ:sum_alpha_1}) there may be some double
counting so that the actual covering probability under the model may
exceed the specified $\alpha$. In the extreme case the actual covering
probability could be as high as $(4+\alpha)/5$ but it will generally
be much lower than this. As an example for a $\alpha=0.9$ and
$\theta=(\exp(3.6),0.3,10)$ 3000 simulations indicate a covering
probability of approximately $\alpha^*=0.94$. If a more or less exact covering
probability is required this can be obtained by replacing $\alpha=0.9$
by ${\tilde \alpha}=2\alpha-\alpha^*=0.86$. Simulations show that the
covering probability is now close to 0.9 as required.

\subsection{Approximating the data}
Given a data set $y(t)$ of size $n$ and the
function $f_{(n,r,\sigma)}$ an approximation interval
$[\varphi_l,\varphi_u]$ for $\varphi$ can be calculated derived as for
(\ref{equ:varphi_approx_4}). A grid size $m_g$ is placed over this
interval which depends on the length of the interval. More precisely
the gird size used is  given by 
\[ng=\max(10, (\varphi_u-\varphi_l)/3)\,.\]
For the given $\alpha$ and $(r,\sigma)$ and a value
of $\varphi$ on the grid simulations are run to calculate the relevant
quantiles of the five statistics used. If all the relevant inequalities
(\ref{equ:inequ_1}) are satisfied with ${\mathbold y}_n$ in place of
${\mathbold Y}_{n,\theta}$, then $P_{\theta}$ with
$\theta=(r,\sigma,\varphi)$ is an adequate approximation to the data.

The output gives the bounds and the values for the data, the five
$p$-values and the minimum $p$-value. For a simulated data of size
$n=100$ with $\theta=(\exp(3.6),0.3,10)$ part of the output with
$\alpha=0.86$ (see above) is given in (\ref{equ:apprx_0}). 
{\footnotesize
\begin{equation} \label{equ:apprx_0}
\left\{
\begin{array}{ccc}
(3.30,0.70,9.62),&(-0.12,2.16,2.38,0.10),&(0.57,1.00,1.05,0.14),\\
(-0.74,-0.57,-0.15,0.40),&(1.49,1.78,2.51,0.28),&(0.07,0.18,0.56).\\
\end{array}
\right.
\end{equation}
}
The first three numbers of (\ref{equ:apprx_0} are the values of $\log
r$, $\sigma$ and $\varphi$ in that order. The following four sets
give, in order, the $(1-\alpha)/10$-quantile of the intercept term for
the three regression coefficients and the residual standard deviation,
followed by the value for the data $y(t)$ followed
by the $(9+\alpha)/10$-quantile followed by the $p$-value. The last
set of three numbers give the Kolmogorov distance for the data, the
$(4+\alpha)/5$-quantile of the Kolmogorov distance and the $p$-value. 
The minimum $p$-value in (\ref{equ:apprx_0}) is 0.1.

A parameter $\theta=(r,\sigma,\varphi)$ is to be judged adequate if
all the empirical values for the regression coefficients lie between the
corresponding order statistics and the the Kolmogorov distance is less
than the order statistic. This is the case for (\ref{equ:apprx_0})
and so $P_{\theta}$ with $\theta=(\exp(3.30),0.70,9.62)$ is judged to
be an adequate approximation to the data. 

\begin{figure}
\begin{center}
\includegraphics[width=4cm,height=12cm,angle=270]{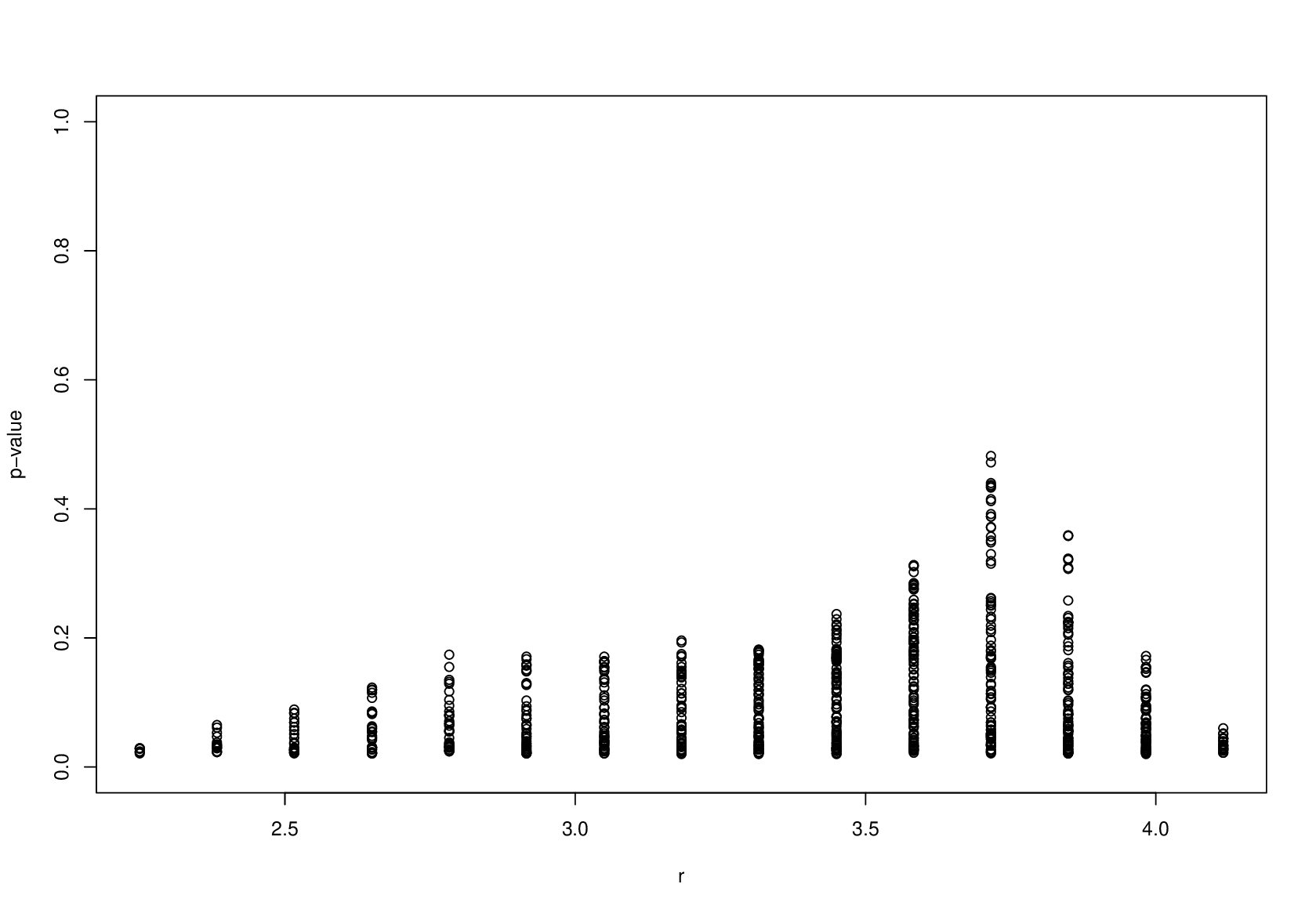}
\includegraphics[width=4cm,height=12cm,angle=270]{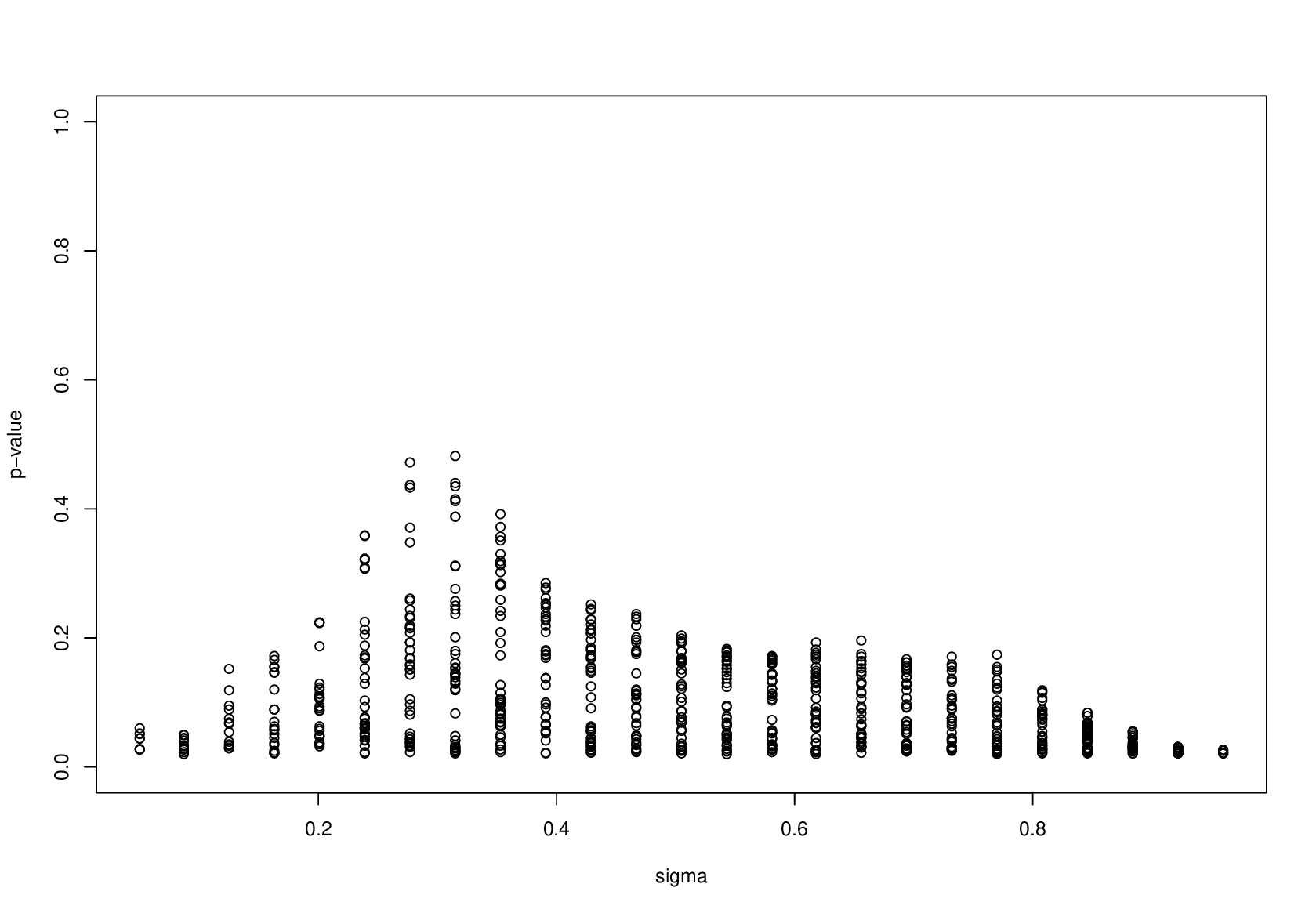}
\includegraphics[width=4cm,height=12cm,angle=270]{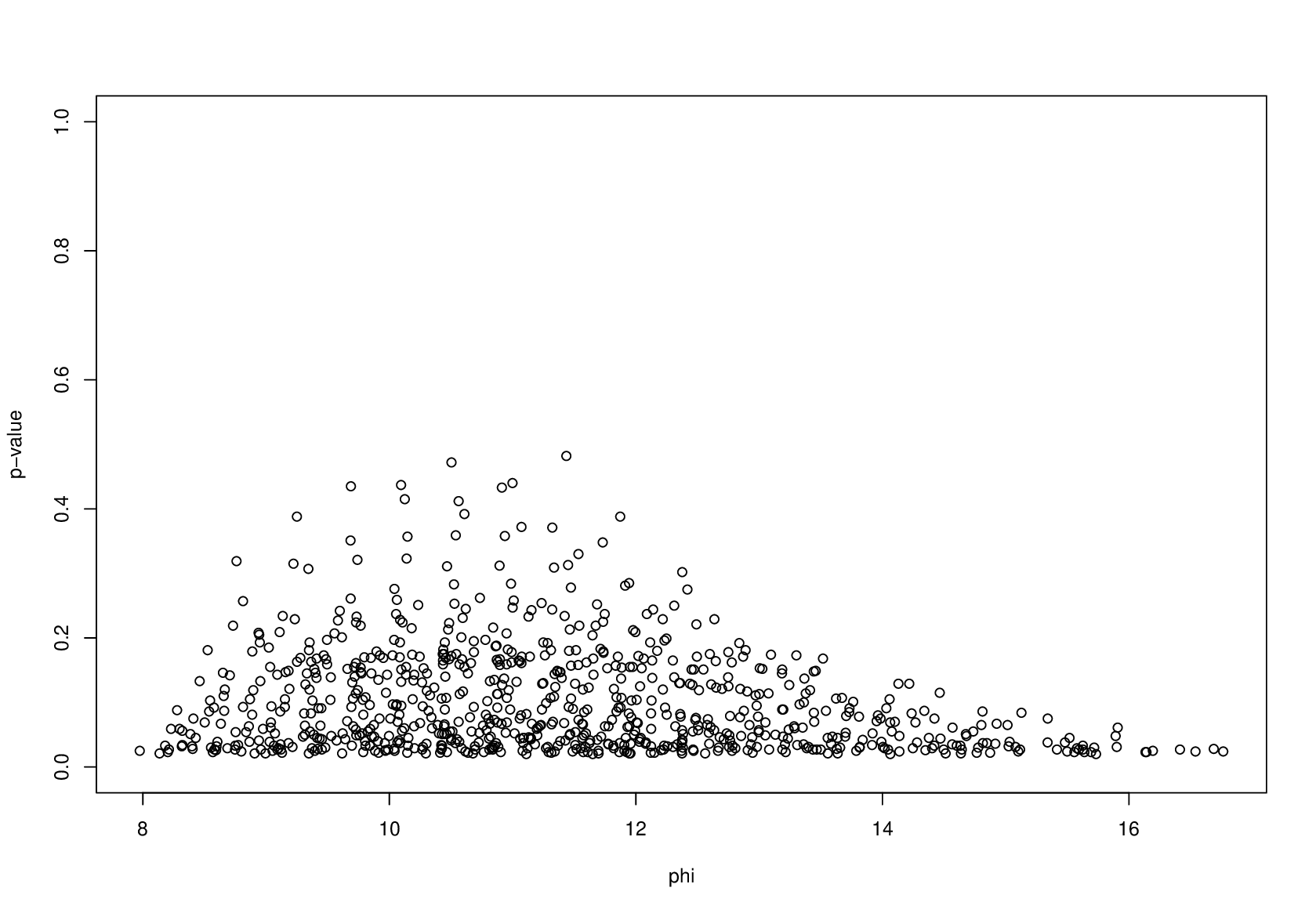}
\end{center}
\caption{Top panel: the minimum $p$-value plotted against $r$. Centre
  panel: the minimum $p$-value plotted against $\sigma$. Bottom panel:
  the minimum $p$-value plotted against $\varphi$. 
\label{fig:p_plot}} 
\end{figure}

For each value of the parameter $\theta=(r,\sigma,\varphi)$ the output
gives 19 values which can be used to asses the degree of
approximation. Likelihood gives only one. Figure~\ref{fig:p_plot}
shows the minimum of the $p$-values plotted against the $r$, $\sigma$
and $\varphi$. This makes sense if all five statistics are treated
equally. The largest minimum value is 0.494 attained for the parameter
constellation $\theta=(\exp(3.716),0.315, 10.97)$. The $\theta$ used to
generate the data was $(\exp(3.6),0.3,10)$.

\begin{figure}
\begin{center}
\includegraphics[width=4cm,height=12cm,angle=270]{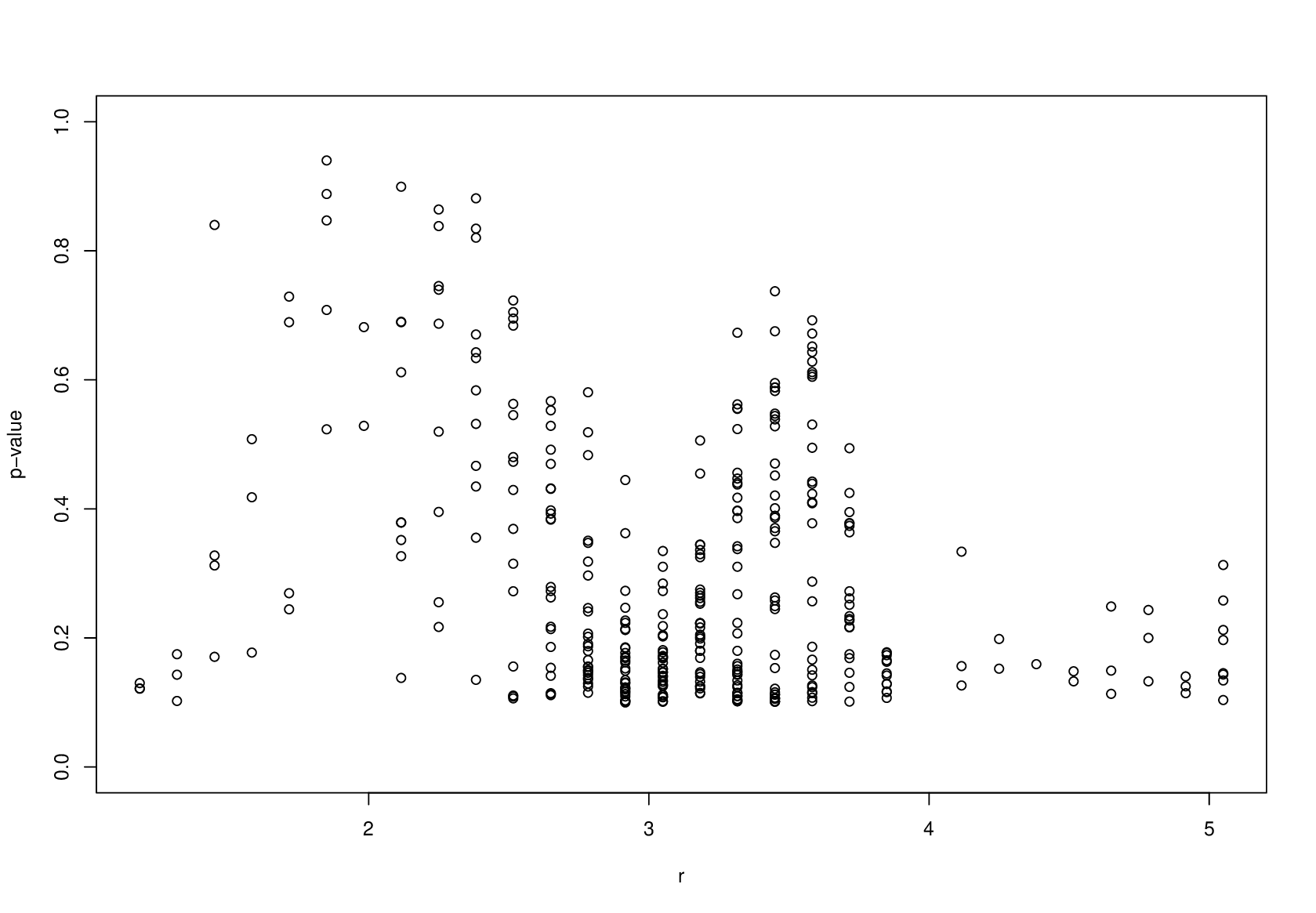}
\includegraphics[width=4cm,height=12cm,angle=270]{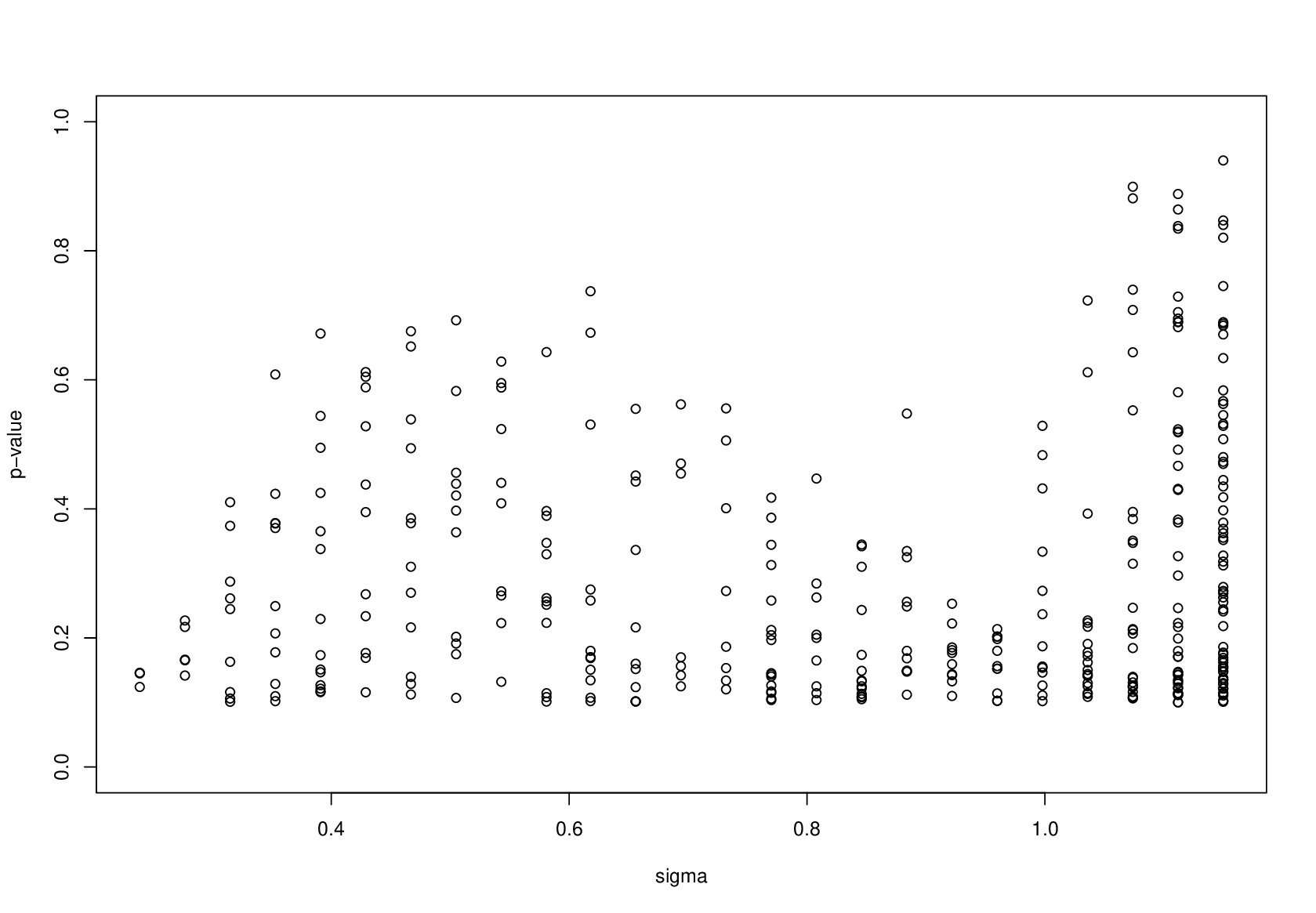}
\includegraphics[width=4cm,height=12cm,angle=270]{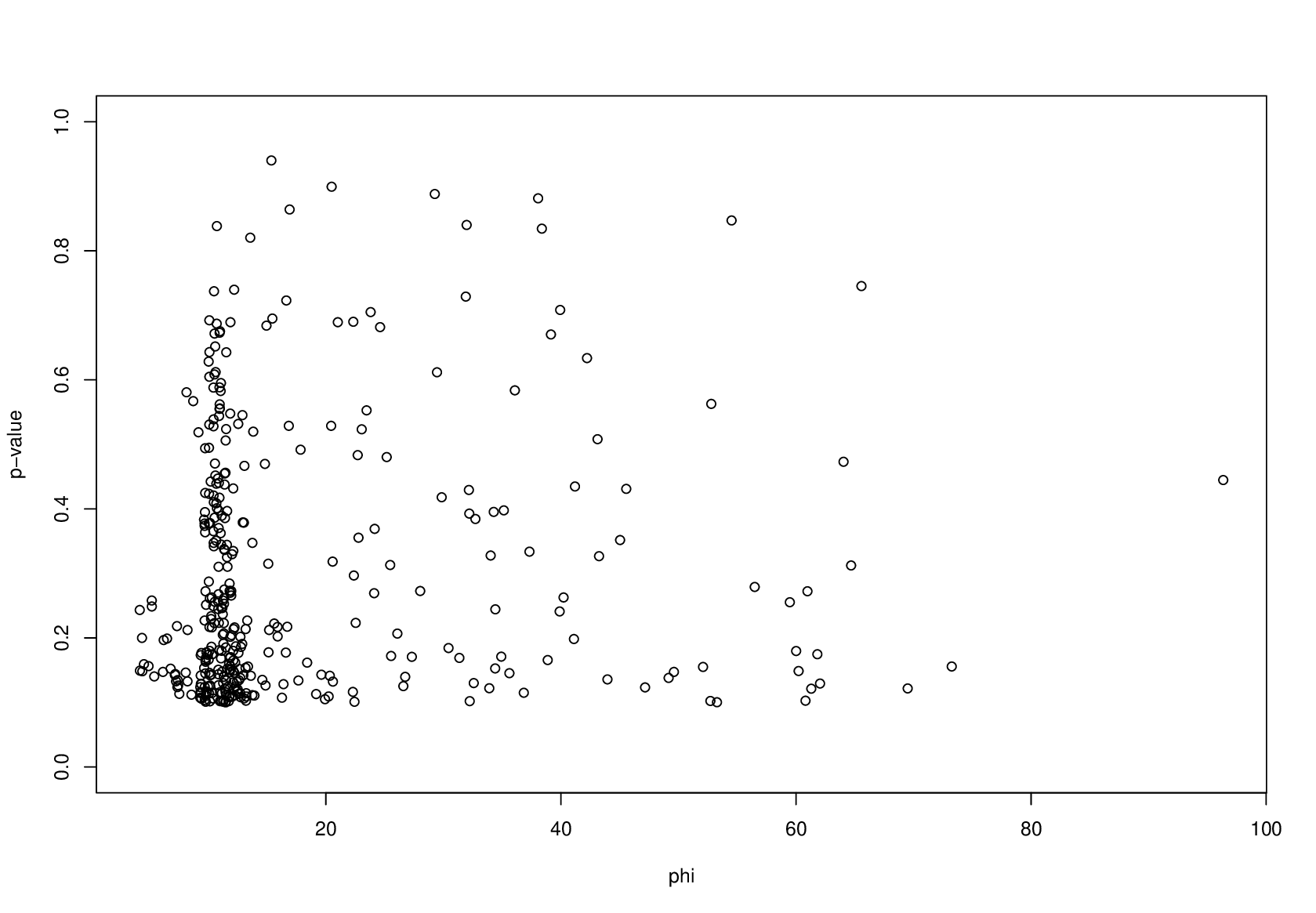}
\end{center}
\caption{Same as for Figure~\ref{fig:p_plot} but with $p$-values based
  on the Mahalanobis distances derived from the covariance
  matrix. \label{fig:p_plot_cv}} 
\end{figure}

If $0.02$ probability is spent on each of the five statistics and if
they were independent then the covering probability would be
$0.98^5=0.904$, that is, very close to the $0.9$ used in the
definition of the approximation region. In fact the four statistics
deriving from the regression  are highly correlated with the
Kolmogorov distance being essentially independent of them. The
correlation can be taken into account by calculating the covariance
matrix and the Mahalanobis distances and then replacing the five
$p$-values by the one based on these distances. Such a single
$p$-value guarantees a correct covering probability. The plots
corresponding to those of Figure~\ref{fig:p_plot} are shown in
Figure~\ref{fig:p_plot_cv} and are very disappointing. The smallest
value of $r$ which is consistent with the data in the sense of the 
Mahalanobis distance is $\exp(1.18)$  with $\theta=(\exp(1.18),1.15,
32.6)$ and a minimum $p$-value of 0.130. 

The reason for this can be
found by analysing the individual simulations. Most of the data sets
of size $n=100$ generated with $\theta=(\exp(1.18),1.15,32.6)$ are so
to speak unexceptional but the occasional one, 0.1\%, may have as many
as 97 zeros. That such data sets occur is due to the high value of
$\sigma=1.15$. They results very high values for the regression
coefficients and these in turn have a considerable influence on the
covariance matrix which has a breakdown point of $1/n$. 

\begin{figure}
\begin{center}
\includegraphics[width=4cm,height=12cm,angle=270]{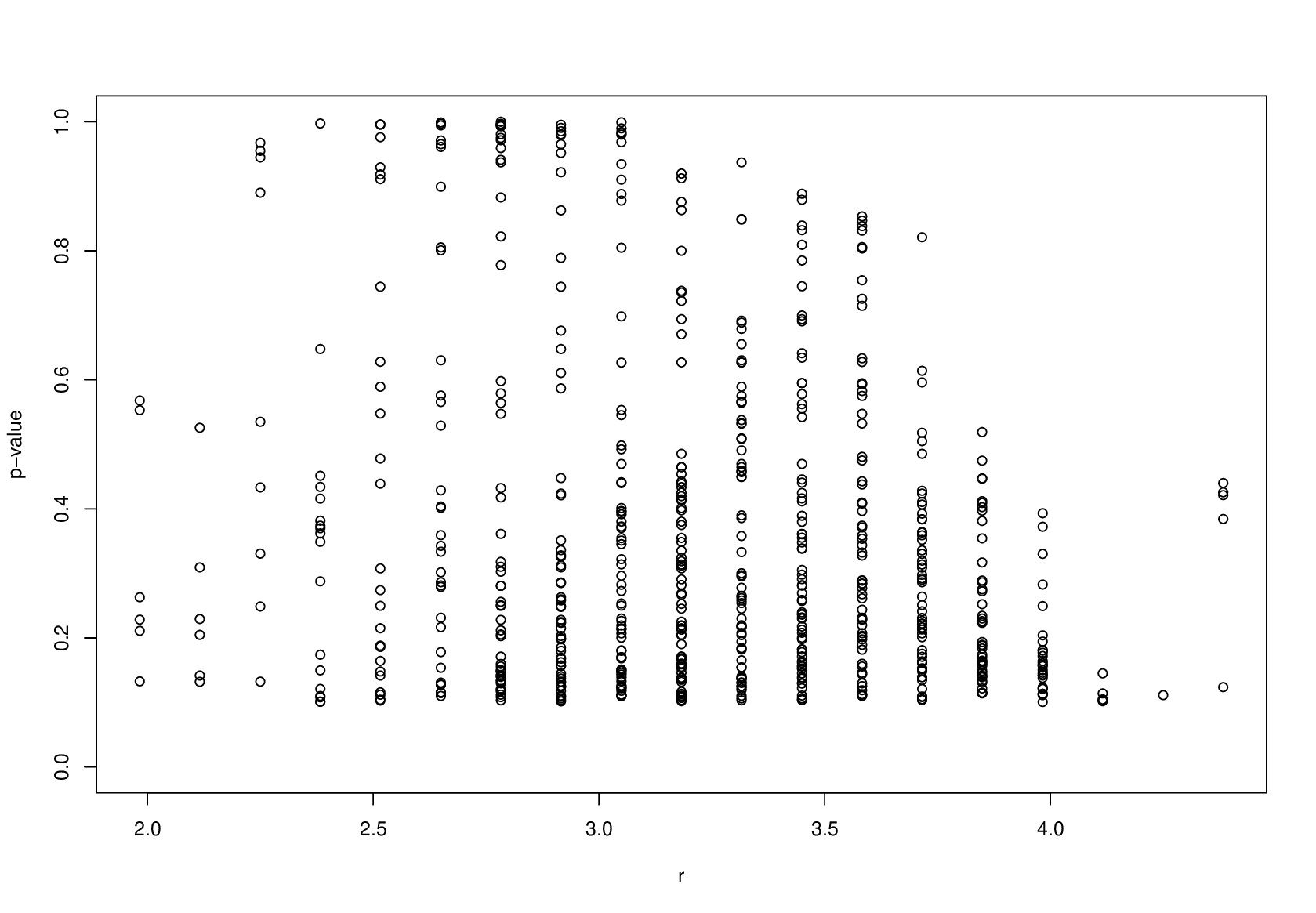}
\includegraphics[width=4cm,height=12cm,angle=270]{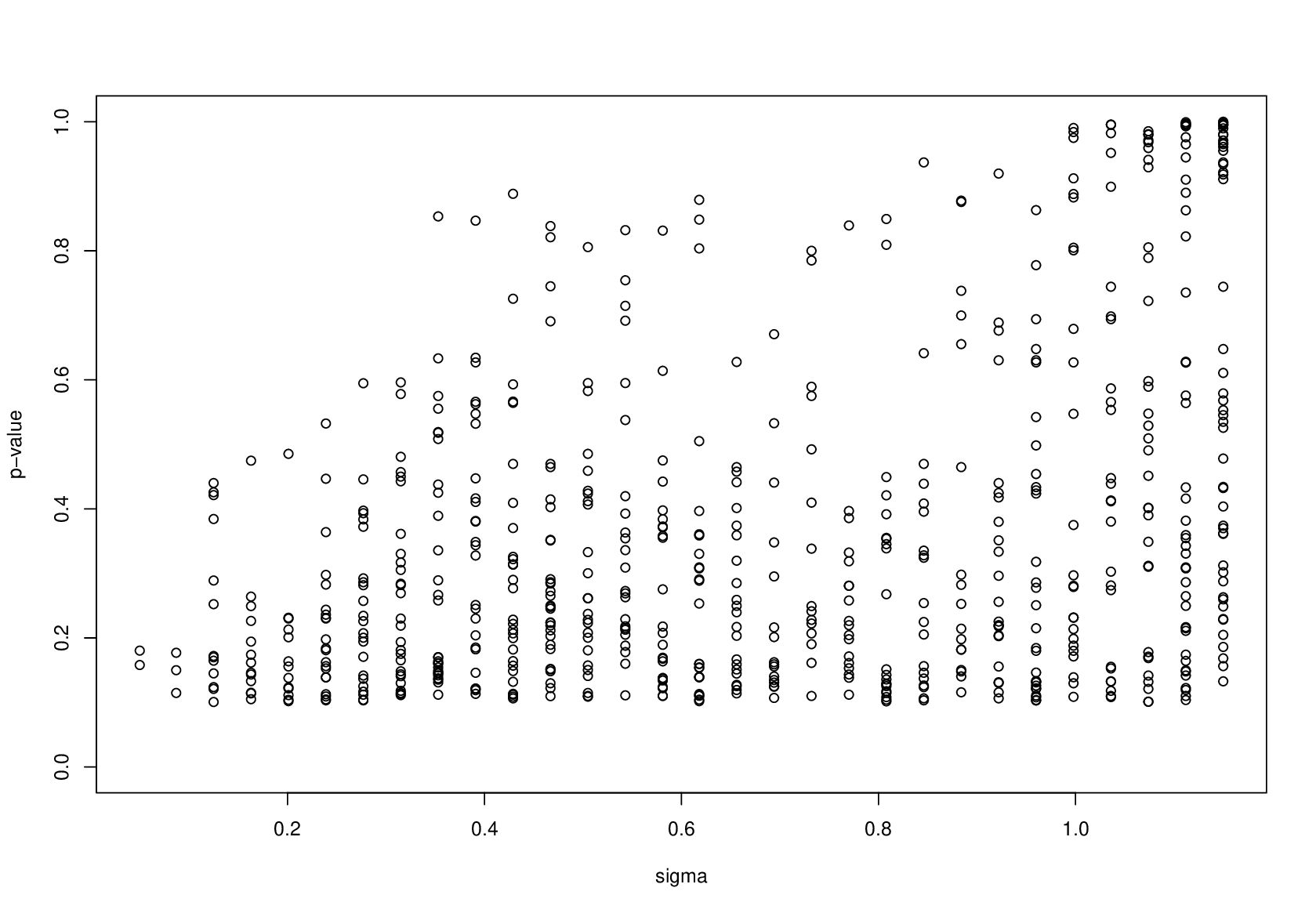}
\includegraphics[width=4cm,height=12cm,angle=270]{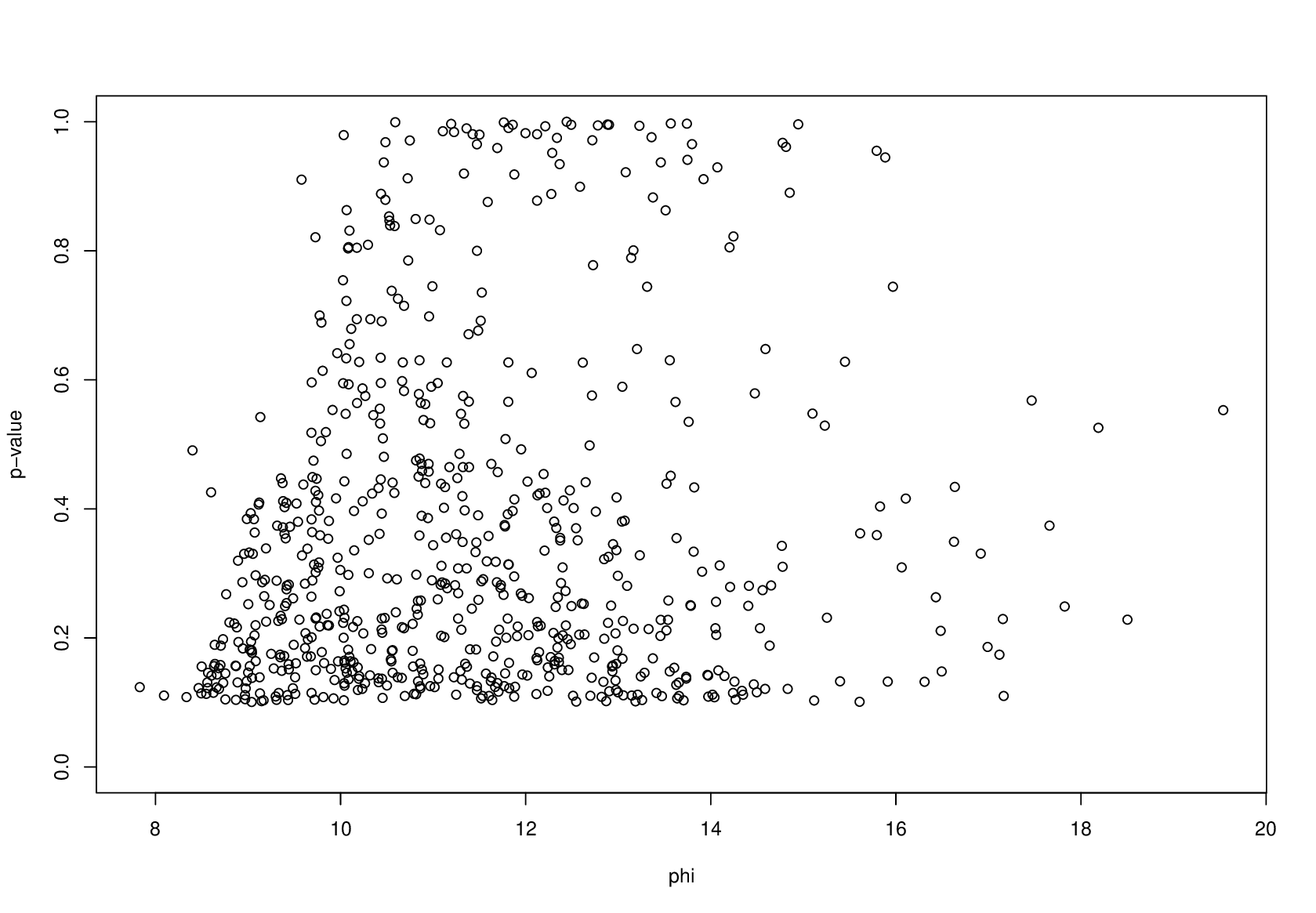}
\end{center}
\caption{Same as for Figure~\ref{fig:p_plot} but with $p$-values based
  Mahalanobis distances derived from the  Kent-Tyler covariance
  matrix with $\nu=2$. 
\label{fig:p_plot_kt}} 
\end{figure}

This last observation suggests using a covariance functional with a
higher breakdown point. The one used here is due to John Kent and
David Tyler and based on the multivariate $t$ distribution
(see \cite{KENTYL91}).  If the $t$ distribution with $\nu$ degrees of
freedom is used the breakdown point in $k$ dimensions is $1/(k+\nu)$
(\cite{DUEMTYL05}). For $\nu=2$ and $k=5$ it is
0.14. Figure~\ref{fig:p_plot_kt} shows the resulting plots. They are
an improvement on Figure~\ref{fig:p_plot_cv} but still disappointing
and much worse than Figure~\ref{fig:p_plot}. It is not clear why this
should be so. One possible reason is that the $p$-values leading to
Figure~\ref{fig:p_plot} are based purely on quantiles and require no
means or covariance matrices. This gives them a robustness not
attainable from covariance functions. If one is only interested in the
best approximation in some sense then attaining the exact covering
probability $\alpha$ is not of high importance. 

The time required for analysing a data set of size $n=100$ is approximately
30 minutes using 3000 simulations and a grid $m_g=10$. If only 1000
simulations are used it reduces to about ten minutes. It is possible
reduce this further by including a rough estimate of
the number of zeros but this will not be discussed further.

\subsection{Tables} \label{sec:tables}
The procedure described above requires values for
$f_{n,r,\sigma}$. These can be obtained for a grid of values for
$(r,\sigma)$ using simulations. The grid taken for the examples in the
paper is given by
\begin{equation}
r=1.05+4i/n_g, \quad \sigma=0.05+1.1(i-1)/(n_g-1), \quad i=1,\ldots,n_g
\end{equation}
with $n_g=30$. Clearly other choices are possible. 
The mean of the order statistics of $N(t)$ is obtained by simulations
for each $(r,\sigma)$ in the grid. In this paper 10000 simulations
were performed for each point $(r,\sigma)$ in the grid. The time
required for the sample size $n=500$ was 35 minutes using Fortran
77 code. The simulations need only be performed once and can be stored for
future use.

\begin{figure}
\begin{center}
\includegraphics[width=4cm,height=12cm,angle=270]{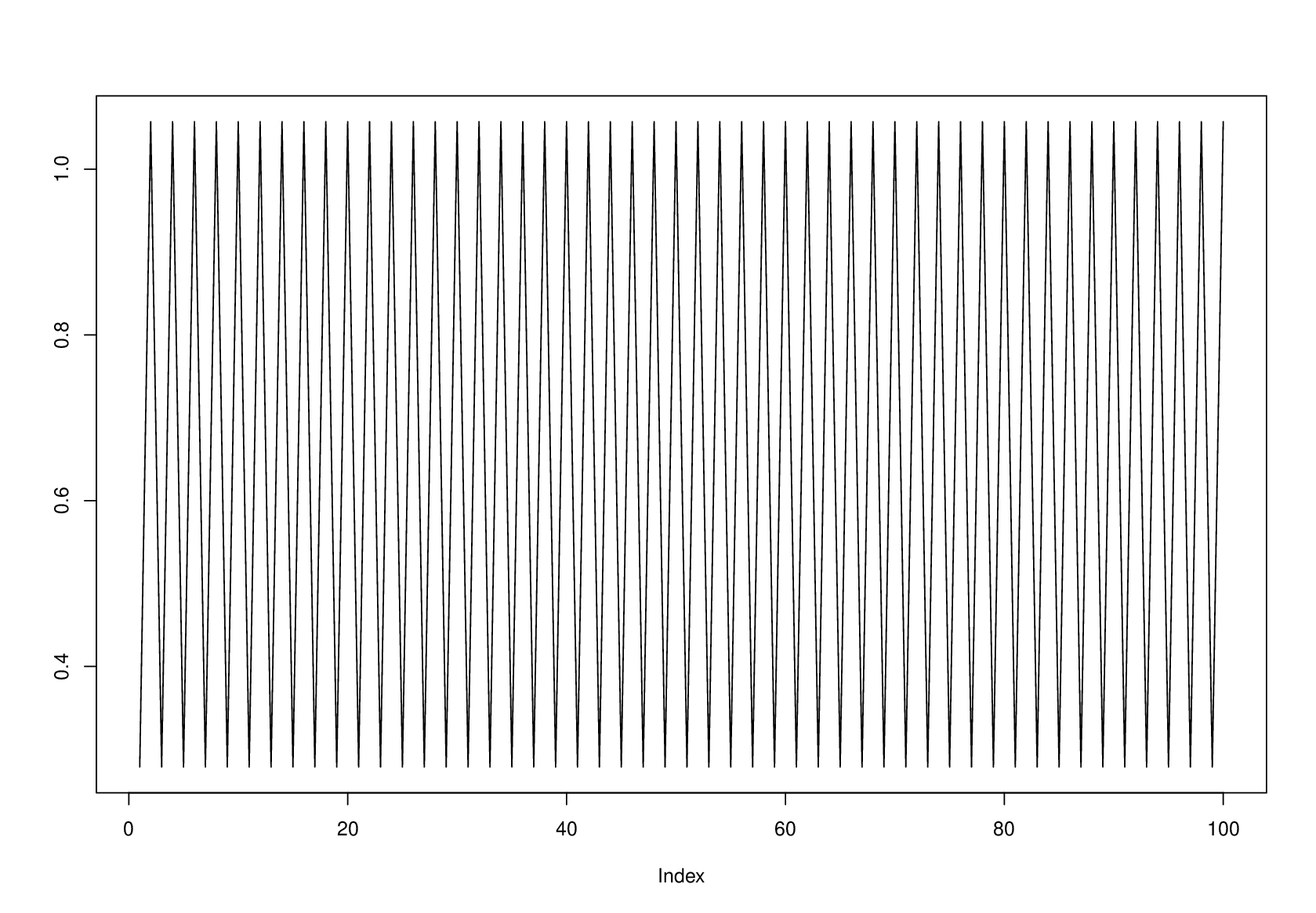}
\includegraphics[width=4cm,height=12cm,angle=270]{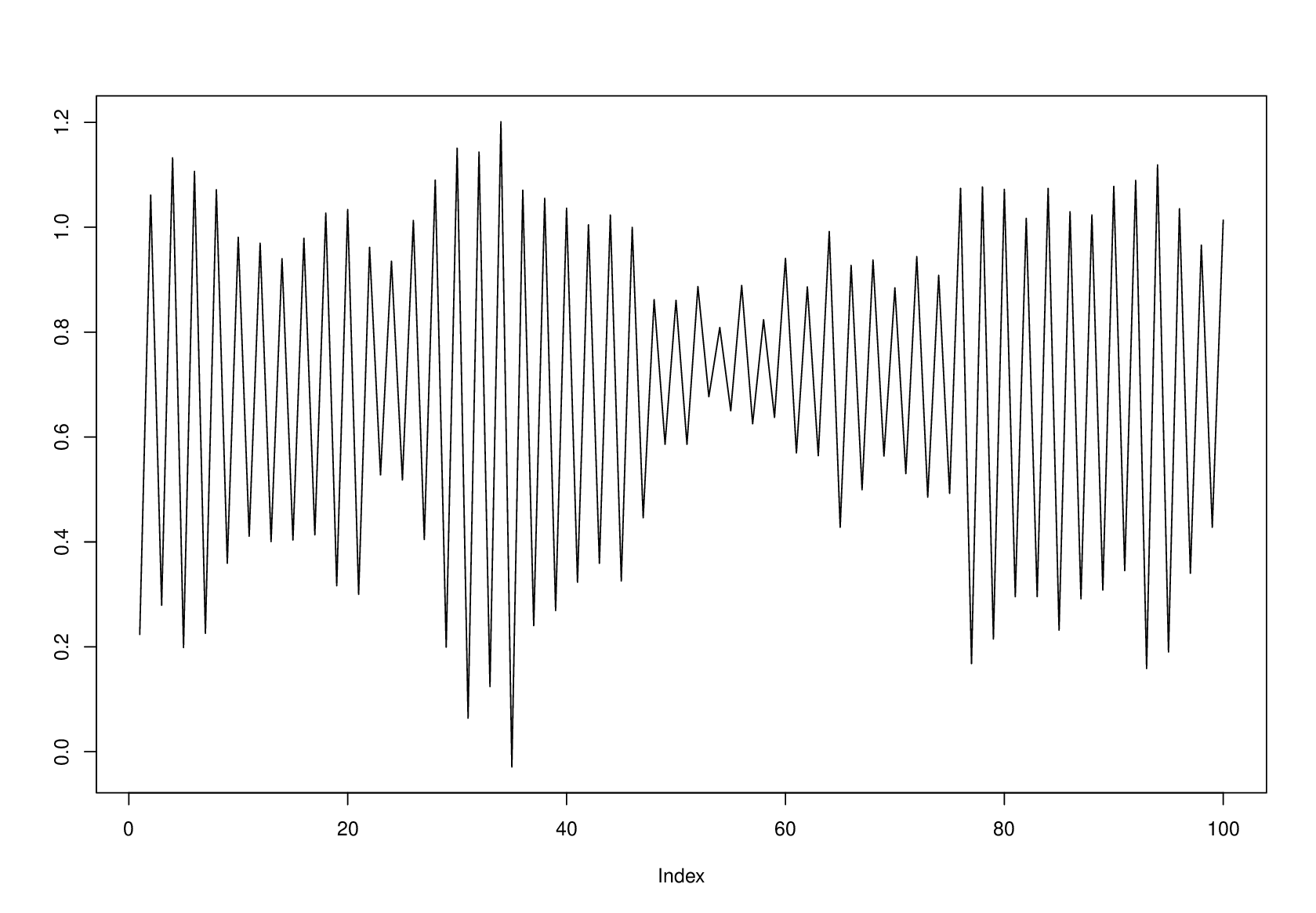}
\includegraphics[width=4cm,height=12cm,angle=270]{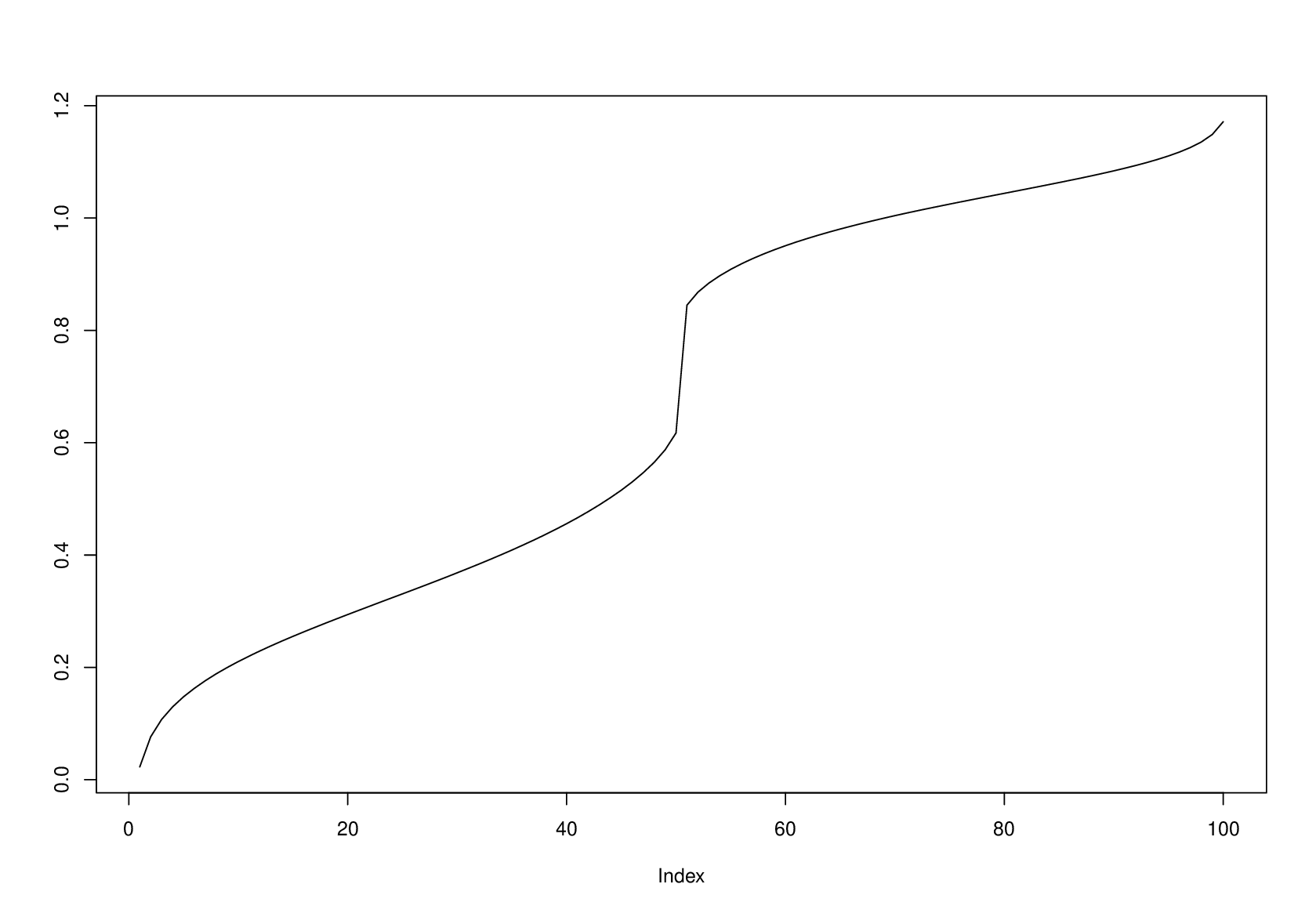}
\end{center}
\caption{Top panel: data set of size 100 of $N(t)$ with
  $(r,\sigma)=(\exp(2.1),0)$. Centre panel:  data set of size 100 of $N(t)$ with
  $(r,\sigma)=(\exp(2.1),0.05)$. Bottom panel: expected values of the
  order statistics of $N(t)$ with $(r,\sigma)=(\exp(2.1),0.05)$.
\label{fig:seven}} 
\end{figure}

The top panel of Figure~\ref{fig:seven} shows data set of size 100 of
$N(t)$ generated with $(r,\sigma)=(\exp(2.1),0)$, that is, a
deterministic Ricker process. The centre panel shows
the same but with  $(r,\sigma)=(\exp(2.1),0.05)$. The bottom panel
shows the expected values of the orders statistics of $N(t)$ again
with  $(r,\sigma)=(\exp(2.1),0.05)$. 
 The bottom panels shows that is is not a simple matter to obtain a
 good and yet parsimonious parametric approximation over the whole
 range.  For this reason 
separate approximations were derived for the first $n/2$ order
statistics and for those for $n/2+1$ to $n-2$. It proved difficult to
approximate the two largest order statistics and so their values were
simply included.
 
A linear regression was used with regressors
\[1,\,x,\,x^2,\,x^3,\, x^4,\,x^5,\,x^6,\,\sin(3\pi x),\,\cos(3\pi x)\]
with $x=1:(n/2)$ for the first $n/2$ order statistics and
$x=(n/2+1):(n-2)$ for the order statistics $n/2+1$ to $n-2$. The
largest error over all the grid points was 3.08 for the order statistics
$1:(n/2)$ and 0.37 for the order statistics $(n/2+1):(n-2)$. the 3.7
may seem large but the values of the order statistics where it occurs
are -20 and less so the 3.08 hardly matter when the exponential is
taken. It is possible that better results can be obtained by splines
but this has not been attempted. The approximation requires storing
9+9+2=20 values for each pair $(r,\sigma)$. Two further values are
required for the approximation interval for $\varphi$, namely the
0.005 and 0.995 quantiles of $\sum_{t=1}^nN(t)$, giving in all 22 values.

\bibliographystyle{natbib}
\bibliography{literature}
\end{document}